\documentclass[12pt]{article}

\usepackage[a4paper,margin=1in]{geometry}
\usepackage{setspace}
\usepackage[T1]{fontenc}
\usepackage[utf8]{inputenc}
\usepackage{lmodern}
\usepackage{microtype}

\usepackage{amsmath,amssymb,amsthm}
\usepackage{mathtools}
\usepackage{bbm}
\usepackage{bm}
\usepackage{commath}
\usepackage{nicefrac}
\usepackage{xfrac}
\usepackage{mathrsfs}

\DeclareMathAlphabet{\mathscr}{OMS}{rsfs}{m}{n}

\theoremstyle{plain}
\newtheorem{theorem}{Theorem}[section]

\theoremstyle{definition}
\newtheorem{definition}[theorem]{Definition}

\theoremstyle{remark}

\newtheorem{example}[theorem]{Example}

\usepackage{graphicx}
\usepackage{caption}
\usepackage{subcaption}
\usepackage{booktabs}
\usepackage{multirow}
\usepackage{array}
\usepackage{float}
\usepackage{tabularx}

\renewcommand{\arraystretch}{1.0}

\usepackage[authoryear]{natbib}

\usepackage{hyperref}
\hypersetup{
    colorlinks=true,
    linkcolor=blue,
    citecolor=blue,
    urlcolor=blue
}

\usepackage{tikz}
\usetikzlibrary{
    decorations.pathmorphing,
    positioning,
    arrows,
    decorations.pathreplacing,
    calc
}

\usepackage{pgfplots}
\usepgfplotslibrary{fillbetween}
\pgfplotsset{compat=1.17}

\usepackage{xcolor}
\usepackage{tcolorbox}

\newtcolorbox{dictionarybox}{
  width=0.9\linewidth,
  center,
  colback=gray!5,
  colframe=gray!60,
  boxrule=0.5pt,
  arc=2pt,
  left=6pt,
  right=6pt,
  top=6pt,
  bottom=6pt,
  title=\textbf{The Calculus--Statistics Dictionary},
  before upper={\renewcommand{\arraystretch}{0.8}}
}

\usepackage{enumitem}
\usepackage{xspace}
\usepackage{comment}
\allowdisplaybreaks

\newcommand{\calM}{\mathcal{M}}
\newcommand{\calT}{\mathcal{T}}

\begin{document}

\begingroup
\setstretch{1.2}

\title{\bf Semiparametric Efficiency Theory as Differential Calculus on a Space of Probability Distributions}
\author{Razieh Nabi\\[0.2em]
Department of Biostatistics and Bioinformatics, Emory University \\[0.2em] 
\texttt{razieh.nabi@emory.edu}}

\date{\today}

\maketitle


\begin{abstract}
Semiparametric efficiency theory provides the mathematical foundation for influence-function-based estimation, including one-step estimators, targeted minimum loss estimators, and many modern inferential methods used in causal inference and missing data analysis. Despite its widespread use, the theory is often presented through a collection of technical constructions whose geometric meaning remains opaque. As a result, influence functions are often derived and applied without an intuitive understanding of the principles connecting scores, tangent spaces, nuisance tangent spaces, and efficient influence functions. This tutorial develops a geometric exposition of semiparametric efficiency theory as a form of differential calculus on a space of probability distributions. Drawing systematic parallels with ordinary multivariable calculus, we show that paths of distributions play the role of curves, scores play the role of velocity vectors, influence functions play the role of gradients, and efficient influence functions arise as projected gradients. This perspective provides a unified explanation for several foundational questions, including why perturbation directions are represented by functions, why tangent spaces depend only on the statistical model whereas nuisance tangent spaces depend on the parameter of interest, and why efficient influence functions arise through orthogonal projection. The resulting framework offers a geometric perspective on semiparametric efficiency theory and influence-function-based inference. 
\end{abstract}



\endgroup

\begingroup
\setstretch{1.3}
\clearpage
{\small 
\tableofcontents
}
\clearpage
\endgroup

\section{Introduction}
\label{sec:intro}

Semiparametric efficiency theory has developed through several complementary research traditions spanning statistics, econometrics, missing data analysis, and causal inference. Foundational ideas can be traced to the development of influence functions, differentiable statistical functionals, and local asymptotic theory \citep{hampel1974influence,pfanzagl1982general,lecam1986asymptotic}. These ideas were subsequently incorporated into a geometric framework for semiparametric inference through the study of tangent spaces, pathwise differentiability, and efficiency bounds \citep{begun1983information,van1991differentiable,chamberlain1987asymptotic,newey1990semiparametric,newey1994asymptotic}. The resulting theory was developed systematically in several influential monographs, including \citet{bickel1993efficient, tsiatis2006semiparametric,kosorok2008introduction}, and Chapter 25 of \citet{van2000asymptotic}. 

The theory has subsequently become central to modern missing data and causal inference methodology. Seminal work by \citet{robins1994estimation, robins1995semiparametric, scharfstein1999adjusting} demonstrated how semiparametric efficiency theory could be used to construct estimators with desirable robustness and efficiency properties in the presence of missing data and treatment assignment mechanisms. These ideas were further developed in the broader framework of causal inference and longitudinal data analysis \citep{laan2003unified}, and later became a cornerstone of targeted learning and influence-function-based estimation strategies \citep{van2011targeted}. More recently, semiparametric efficiency theory has provided much of the conceptual and mathematical foundation for modern machine-learning-assisted inference, including doubly robust estimation, targeted minimum loss estimation, and debiased machine learning procedures \citep{kennedy2016semiparametric,chernozhukov2018double, hines2022demystifying}.

Despite its maturity and practical importance, semiparametric efficiency theory can present a substantial learning curve. The subject is built around a collection of interconnected concepts, including paths, scores, tangent spaces, nuisance tangent spaces, orthogonal complements, and efficient influence functions. While mathematically elegant, the geometric relationships among these objects are often not immediately apparent, obscuring the intuition behind why these constructions arise and how they fit together.  

Many of the central constructions of semiparametric efficiency theory have direct analogues in ordinary differential calculus. In multivariable calculus, one studies how a function changes as a point moves through Euclidean space. In semiparametric efficiency theory, one studies how a statistical functional changes as a probability distribution moves through a statistical model. Paths play the role of curves, scores play the role of velocity vectors, influence functions play the role of gradients, and efficient influence functions arise as projected gradients. Much of semiparametric efficiency theory may therefore be viewed as a form of differential calculus carried out on a space of probability distributions. We adopt this perspective as the organizing principle of the tutorial.

The perspective developed here is not new. The ideas are deeply embedded in the foundations of semiparametric inference and appear throughout the literature in various forms. The contribution of this paper is therefore not a new theoretical framework, but a unifying exposition. Many of the geometric insights underlying semiparametric inference are dispersed across different sources and introduced at different levels of abstraction. Our goal is to organize these ideas into a single conceptual narrative built around a differential-calculus perspective. By making the underlying geometry explicit, we aim to clarify both the meaning of the central constructions and the relationships among them.

This observation motivates the central theme of the tutorial. Rather than introducing the main objects of semiparametric efficiency theory independently, we develop them as natural answers to a single organizing question:
\begin{center}
\emph{How do we perform differential calculus when the object being differentiated is a probability distribution rather than a point in  Euclidean space?}
\end{center}

Viewed through this lens, several questions that often appear technically unrelated become part of a common geometric narrative. Some concern the nature of movement itself: what exactly is moving when a probability distribution is perturbed, and why are perturbation directions represented by functions rather than vectors? Others concern the distinction between possible and relevant directions: why does the tangent space depend only on the statistical model, whereas the nuisance tangent space depends on the parameter of interest? Finally, once directions have been identified, how should the resulting derivative be represented, and why do orthogonality, projection, and influence functions arise naturally in that representation?

A recurring theme is that the local geometry is governed by two distinct structures. The statistical model determines the directions in which a distribution is allowed to move, while the parameter determines how changes in those directions are measured. Many of the central constructions of semiparametric efficiency theory emerge from the interaction between these two structures. 

Our goal is conceptual clarity rather than technical generality. We assume familiarity with basic probability and statistical inference but no prior exposure to semiparametric efficiency theory. The correspondences with
ordinary multivariable calculus are summarized in the \emph{Calculus--Statistics Dictionary} in Appendix~\ref{app:sec:notation}. We also show how this geometric viewpoint connects influence functions to statistical estimation, including one-step estimators, targeted minimum loss estimators, and other modern semiparametric procedures. Table~\ref{tab:notation} in Appendix~\ref{app:sec:notation} also summarizes the principal notation.

The remainder of the paper proceeds as follows. 
Section~\ref{sec:diff_calc_vs_stats} introduces statistical parameters as functionals of probability distributions and explains what it means for a distribution to move. 
Section~\ref{sec:directions_tangent_space} develops scores and tangent spaces as the statistical analogues of velocity vectors and tangent planes. 
Section~\ref{sec:nuisance_directions} introduces nuisance tangent spaces and the distinction between informative and uninformative perturbations. 
Section~\ref{sec:IFs_gradients} develops influence functions as gradients of statistical functionals.  
Section~\ref{sec:EIF} explains efficient influence functions through orthogonality and projection. 
Section~\ref{sec:estimation} connects the geometric theory to statistical estimation.  
Section~\ref{sec:examples} illustrates the framework through familiar examples.  
Section~\ref{sec:conclusion} concludes the paper.

\section{Objects, Functionals, and Paths}
\label{sec:diff_calc_vs_stats}

The starting point of semiparametric efficiency theory is an analogy with ordinary multivariable calculus. In calculus, one studies how a function $f:\mathbb R^d\to\mathbb R$ changes as its argument $x\in\mathbb R^d$ moves through Euclidean space. In semiparametric efficiency theory, one studies how a statistical parameter $\psi(P)$ changes as its argument $P$ moves through a collection of probability distributions, called a statistical model, denoted by $\mathcal M$. The basic correspondence is therefore that points $x$ are replaced by probability distributions $P$, and ordinary functions $f$ are replaced by statistical functionals $\psi$. The purpose of this section is to make this correspondence precise and to clarify what it means for a probability distribution to ``move.''  

Let $O$ denote an observed data vector taking values in a measurable sample space $\mathcal O$. We assume that the data are generated from an unknown probability distribution $P_0$. A statistical model $\mathcal M$ is a collection of distributions regarded as plausible candidates for $P_0$. A parameter of interest is a map $\psi:\mathcal M\to\mathbb R$, which assigns a numerical quantity to each distribution in the model. For example, if $Y$ denotes an outcome variable, the population mean is the functional $\psi(P)=E_P\{Y\}$. Throughout the paper, $P$ denotes a generic distribution in the model, while $P_0$ denotes the true data-generating distribution. The quantity of ultimate interest is therefore $\psi(P_0)$, the value of the parameter at the true distribution. 


The statistical model plays a role analogous to the domain on which a function is defined. Statistical models are often classified according to the structure used to index their distributions. A \emph{parametric model} is indexed by a finite-dimensional parameter, $\mathcal M = \{P_\theta:\theta\in\Theta\subseteq\mathbb R^k\}$. A \emph{semiparametric model} is typically indexed by a finite-dimensional parameter $\theta$ together with an infinite-dimensional nuisance parameter $\eta$ and may be represented as $\mathcal M=\{P_{\theta,\eta}:\theta\in\Theta\subseteq\mathbb R^k,\, \eta\in\mathcal H\}$, where $\mathcal H$ is an infinite-dimensional parameter space. A \emph{nonparametric model} does not impose a finite-dimensional parameterization on the data-generating distribution
and is itself infinite-dimensional. 

The geometric viewpoint developed here does not require the statistical model itself to admit a global parameterization by $(\theta,\eta)$. Indeed, semiparametric efficiency theory is routinely used to study finite-dimensional functionals $\psi(P)$ of an otherwise nonparametric distribution $P$. In this setting, the term \emph{semiparametric} refers to the inferential problem of estimating a finite-dimensional target in the presence of an infinite-dimensional
nuisance structure. What matters for the local theory is the collection of perturbations of $P_0$ that remain inside the statistical model. Restricting $\mathcal M$ restricts these allowable directions of movement. Thus, the geometry relevant to efficiency is determined locally by the model around $P_0$, a point we make precise through paths and tangent spaces.

Semiparametric efficiency theory is ultimately concerned with understanding how well statistical parameters can be estimated and how estimators should be constructed. A central idea is that these questions can be studied by examining how a parameter responds to small perturbations of the underlying distribution. Before discussing derivatives, gradients, influence functions, or efficiency, we must therefore answer a more basic question: what exactly is changing?

The discussion that follows involves four related objects: $\mathcal O,\, O,\, o,\, P$. Here, $\mathcal O$ denotes the sample space, $O$ denotes the observed data vector, $o$ denotes a possible realization of that data vector, and $P$ denotes a probability distribution on $\mathcal O$. These objects play different roles. Throughout semiparametric efficiency theory, we consider perturbations that change the probability distribution while leaving the underlying data structure unchanged. The sample space $\mathcal O$, the random variable $O$, and the set of possible realizations $o \in \mathcal O$ remain the same. What changes is the probability distribution $P$, that is, the probabilities assigned to the possible observations. 

This distinction is conceptually important because geometric language can otherwise suggest that the observations themselves are moving. They are not. A perturbation of a statistical model changes the probabilities assigned to possible observations, not the observations themselves. Semiparametric efficiency theory is therefore concerned with local perturbations of the data-generating distribution while holding fixed the underlying sample space and data structure. In this sense, the analogue of a moving point in ordinary calculus is not a moving observation, but a moving distribution. 

Once probability distributions are recognized as the objects that move, we need a way to describe local perturbations of the statistical model. In ordinary multivariable calculus, given a point $x_0\in\mathbb R^d$ and a direction $v\in\mathbb R^d$, one may consider the line $x_\varepsilon = x_0+\varepsilon v$, so that $x_{\varepsilon=0}=x_0$. The parameter $\varepsilon$ indexes movement away from the reference point $x_0$, and the behavior of $f(x_\varepsilon)$ reveals how the function changes around $x_0$ in direction $v$. Semiparametric efficiency theory follows the same idea. Instead of a point $x_0$, we begin with a probability distribution $P_0$. Instead of a line of points, we consider a family of probability distributions indexed by $\varepsilon$ and passing through $P_0$. To develop a differential theory, such paths must be differentiable at $P_0$ in a sense that permits a first-order local approximation. The relevant notion of smoothness will be introduced in the next section through differentiability in quadratic mean, which is tied to the Hellinger geometry of probability distributions. For now, we focus only on the basic idea of a \textit{path} as a one-dimensional perturbation through the statistical model.

Formally, let $P_0 \in \mathcal M$ denote the true data-generating distribution. A path through $P_0$ is a collection of distributions
\[
\{P_\varepsilon : \varepsilon \in (-\delta,\delta)\}
\subseteq
\mathcal M
\]
such that $P_{\varepsilon=0}=P_0$. The parameter $\varepsilon$ indexes local movement away from the reference distribution, while the requirement $P_\varepsilon \in \mathcal M$ ensures that the perturbation remains inside the statistical model. As in ordinary differential geometry, all local constructions that follow—including scores and tangent spaces—will be defined relative to the reference distribution $P_0$.

To develop a differential theory, we restrict attention to paths that admit a well-defined first-order local approximation. The standard smoothness condition is differentiability in quadratic mean at $\varepsilon=0$, introduced formally in Appendix~\ref{app:sec:dqm}. Throughout the paper, we refer to such paths as \emph{regular paths}. In the semiparametric literature, paths of this type are commonly introduced through one-dimensional regular parametric submodels passing through $P_0$. We use the term path to emphasize the geometric analogy with curves in ordinary calculus.


\begin{figure}[t]
\centering

\begin{subfigure}[t]{0.48\textwidth}
\centering

\begin{tikzpicture}[scale=0.9]

\draw[rounded corners=18pt, thick, gray!60]
(-3.2,-1.8) rectangle (3.,1.8);

\node[gray!70] at (-2.5,1.45) {$\mathcal M$};

\draw[thick, ->]
plot[smooth, tension=0.9] coordinates {
(-2.4,-1.4)
(-1.4,-0.65)
(0,0)
(1.35,0.72)
(2.35,1.45)
};

\filldraw (-2.4,-1.4) circle (2pt);
\filldraw (-1.4,-0.65) circle (2pt);
\filldraw (0,0) circle (2.4pt);
\filldraw (1.35,0.72) circle (2pt);
\filldraw (2.35,1.45) circle (2pt);

\node[below left] at (0.2,0.65) {$P_0=P_{\varepsilon=0}$};
\node[below left] at (-0.5,-0.65) {$P_{\varepsilon<0}$};
\node[above right] at (1.3,0.22) {$P_{\varepsilon>0}$};


\end{tikzpicture}
\end{subfigure}
\begin{subfigure}[t]{0.48\textwidth}
\centering

\begin{tikzpicture}[scale=0.85]

\begin{scope}[xshift=-3cm]
\node at (0.65,2.) {$P_{\varepsilon<0}$};

\draw[->] (-0.7,0) -- (1.7,0);

\filldraw[fill=gray!30] (-0.3,0) rectangle (0.3,1.4);
\filldraw[fill=gray!30] (0.7,0) rectangle (1.3,0.6);

\node at (0,-0.3) {$0$};
\node at (1,-0.3) {$1$};
\end{scope}

\begin{scope}
\node at (0.65,2.) {$P_0$};

\draw[->] (-0.7,0) -- (1.7,0);

\filldraw[fill=gray!30] (-0.3,0) rectangle (0.3,1.0);
\filldraw[fill=gray!30] (0.7,0) rectangle (1.3,1.0);

\node at (0,-0.3) {$0$};
\node at (1,-0.3) {$1$};
\end{scope}

\begin{scope}[xshift=3cm]
\node at (0.7,2.) {$P_{\varepsilon>0}$};

\draw[->] (-0.7,0) -- (1.7,0);

\filldraw[fill=gray!30] (-0.3,0) rectangle (0.3,0.6);
\filldraw[fill=gray!30] (0.7,0) rectangle (1.3,1.4);

\node at (0,-0.3) {$0$};
\node at (1,-0.3) {$1$};
\end{scope}


\end{tikzpicture}

\end{subfigure}

\caption{\emph{(left)} A path $\{P_\varepsilon:\varepsilon\in(-\delta,\delta)\}$ through a statistical model $\mathcal M$. The distribution $P_0$ is a reference point, and ${P_\varepsilon}$ traces nearby distributions as $\varepsilon$ varies, while remaining inside $\mathcal M$. \emph{(right)} Movement in a statistical model for a Bernoulli distribution. The possible observations remain $\{0,1\}$, but the probability mass assigned to those observations changes as $\varepsilon$ varies. }
\label{fig:path-statistical-model}
\end{figure}
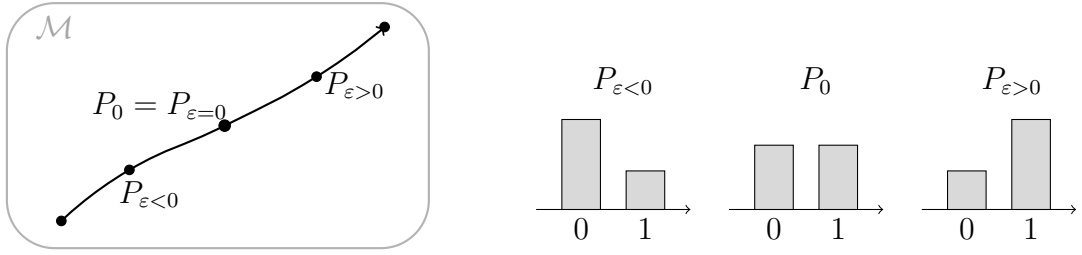

Geometrically, a statistical model may be viewed as a space whose points are probability distributions. A path may therefore be viewed as a curve through the statistical model; see the left panel in Figure~\ref{fig:path-statistical-model} for an illustration. Introducing paths allows us to ask the same local questions that arise in ordinary calculus: How can we describe the infinitesimal movement of a probability distribution? What constitutes its direction of movement? And how does a parameter $\psi(P)$ respond to that movement? These questions lead naturally to the notions of scores, tangent spaces, and pathwise derivatives.

\begin{example}
Suppose $O\sim \mathrm{Bernoulli}(p)$, so that $P(O=1)=p, \, P(O=0)=1-p$. The statistical model is $\mathcal M=\{P_p:0<p<1\}$, where each value of $p$ corresponds to a different probability distribution on the fixed sample space $\{0,1\}$. Let $P_0$ denote the distribution corresponding to $p_0$. In this one-dimensional model, a path through $P_0$ can be constructed by varying the Bernoulli parameter locally: $p_\varepsilon=p_0+\varepsilon a$, where $a$ is fixed and $\varepsilon$ is sufficiently small that $0<p_\varepsilon<1$. The corresponding family $P_\varepsilon=P_{p_\varepsilon}$ is a path through $P_0$.
\end{example}

The above example illustrates precisely what is meant by movement in a statistical model. The sample space remains $\{0,1\}$, the random variable remains Bernoulli, and the possible observations are still $0$ and $1$. What changes are the probabilities assigned to those observations. Under the perturbed distribution, $P_\varepsilon(O=1)=p_\varepsilon$ and $P_\varepsilon(O=0)=1-p_\varepsilon$. As $\varepsilon$ varies, probability mass is redistributed between the two outcomes while the set of possible observations remains unchanged. Thus, movement in a statistical model is movement of probability mass, not movement of observations. See the right panel in Figure~\ref{fig:path-statistical-model} for an illustration. 

The Bernoulli example is finite-dimensional, but it captures the key geometric idea. A statistical model is a collection of probability distributions, and a path describes a local way of moving through that collection. In the next section, we ask a more refined question: once a path has been specified, how do we describe the direction in which the distribution is moving? In ordinary calculus, this information is captured by a velocity vector. In semiparametric efficiency theory, the corresponding object is the score function. 

\section{Directions and Tangent Spaces}
\label{sec:directions_tangent_space}

In the previous section, we introduced paths of probability distributions as the statistical analogue of curves in ordinary geometry. The next step is to describe the direction in which such a path moves. Scores, tangent spaces, nuisance tangent spaces, and influence functions all arise from the attempt to formalize what it means for a probability distribution to move infinitesimally within a statistical model.

\subsection{Scores as statistical velocities}

To motivate the statistical construction, first recall the finite-dimensional setting. Let $\epsilon \mapsto x_\epsilon$ be a differentiable curve in $\mathbb R^d$ passing through $x_0$ at $\varepsilon=0$. Its local behavior is summarized by the velocity vector
\[
v
=
\left.
\frac{d}{d\varepsilon}
x_\varepsilon
\right|_{\varepsilon=0}. 
\]

\noindent 
By the first-order Taylor expansion $x_\varepsilon=x_0+\varepsilon v+o(\varepsilon)$, the vector $v$ captures the first-order behavior of the curve near $x_0$. In the same way, we seek a first-order description of a path of distributions $P_\varepsilon$ near $P_0$.


Suppose that the distributions along a regular path $\{P_\varepsilon\}$ admit densities $p_\varepsilon$ with respect to a common dominating measure. Formally, regularity is defined through differentiability in quadratic mean, as described in Appendix~\ref{app:sec:dqm}. For intuition, however, first consider the stronger pointwise setting in which the density path admits the expansion
\[
p_\varepsilon(o)
=
p_0(o)
+
\varepsilon \dot p(o)
+
\mathrm{o}(\varepsilon),
\qquad
\dot p(o)
=
\left.
\frac{\partial}{\partial\varepsilon}
p_\varepsilon(o)
\right|_{\varepsilon=0}.
\]
The function $\dot p(o)$ records the pointwise first-order change in the density or mass function under the perturbation.
The corresponding relative first-order change is
\[
S(o)
=
\frac{\dot p(o)}{p_0(o)}
=
\left.
\frac{\partial}{\partial\varepsilon}
\log p_\varepsilon(o)
\right|_{\varepsilon=0},
\]
where $p_0(o)>0$. The function $S$ is called the \emph{score} associated
with the path. It provides the statistical representation of the
first-order direction of movement of the distribution.

This pointwise construction is useful for intuition but is stronger than the smoothness condition required by the general theory. Formally, differentiability in quadratic mean describes the first-order movement of the square-root density path in $L^2$ geometry and defines the score $S$ directly as an element of $L_0^2(P_0)$. Under the stronger pointwise smoothness conditions used above, the quadratic-mean score agrees with the familiar log-density derivative. We develop this connection formally in Appendix~\ref{app:sec:dqm}. Thus, strictly speaking, the score is a representation of the local velocity of the distribution path under the quadratic-mean geometry. Throughout the paper, we use the shorthand that scores play the role of statistical velocity vectors.

Since $\int p_\varepsilon(o)\,do=1$ for all $\varepsilon$. Differentiating at $\varepsilon=0$ gives $\int \dot p(o)\,do=0$, or equivalently $E_{P_0}\{S(O)\}=0$. Under standard regularity conditions, scores also satisfy $E_{P_0}\{S(O)^2\}<\infty$. Consequently, every score arising from a regular path belongs to the space of all square-integrable mean-zero functions under $P_0$, defined as  

\vspace{-0.5cm}
\[
L_0^2(P_0)
=
\left\{
f:
E_{P_0}\{f(O)\}=0, \, 
E_{P_0}\{f(O)^2\}<\infty
\right\}. 
\]


The $L_0^2(P_0)$ space will provide the natural setting for describing perturbation directions. In particular, it allows us to collect and compare different score directions, which leads naturally to the notion of a tangent space developed next.

\begin{figure}[t]
\centering
\begin{tikzpicture}[scale=1.0]

\draw[->] (0,0) -- (4.2,0) node[right] {$o$};
\draw[->] (0,0) -- (0,2.7) node[above] {$p(o)$};

\draw[thick, smooth] plot coordinates {
(0.3,0.25) (0.8,0.65) (1.3,1.7) (1.8,1.05) (2.4,0.55) (3.2,0.25) (3.9,0.15)
};
\node at (2.1,-0.45) {baseline density $p_0(o)$};

\draw[->, thick] (4.8,1.25) -- (7.0,1.25)
node[midway, above] {\small $\varepsilon > 0 $ perturbation};

\draw[->] (8.6,0) -- (12.8,0) node[right] {$o$};
\draw[->] (8.6,0) -- (8.6,2.7) node[above] {$p(o)$};

\draw[dashed, smooth] plot coordinates {
(8.9,0.25) (9.4,0.65) (9.9,1.7) (10.4,1.05) (11.0,0.55) (11.8,0.25) (12.5,0.15)
};

\draw[thick, smooth] plot coordinates {
(8.9,0.20) (9.4,0.50) (9.9,1.35) (10.4,1.25) (11.0,0.85) (11.8,0.35) (12.5,0.15)
};

\node at (10.7,-0.45) {perturbed density $p_\varepsilon(o)$};

\draw[->] (9.65,2.15) -- (9.9,1.38);
\node[align=center] at (10.75,2.35) {\small $S(o)<0$, mass decreases};

\draw[->] (11.15,1.35) -- (11.0,0.85);
\node[align=center] at (12.65,1.6) {\small $S(o)>0$, mass increases};
\end{tikzpicture}
\caption{A path induces a score. The score describes the local redistribution of probability mass under an infinitesimal perturbation ($p_\varepsilon(o) = p_0(o)\{1+\varepsilon S(o)\} + o(\varepsilon)$). 
Positive values of $S(o)$ identify locations in the sample space where density or probability mass increases to first order under the perturbation, whereas negative values identify locations where it decreases to first order.}
\label{fig:path-score}
\end{figure}
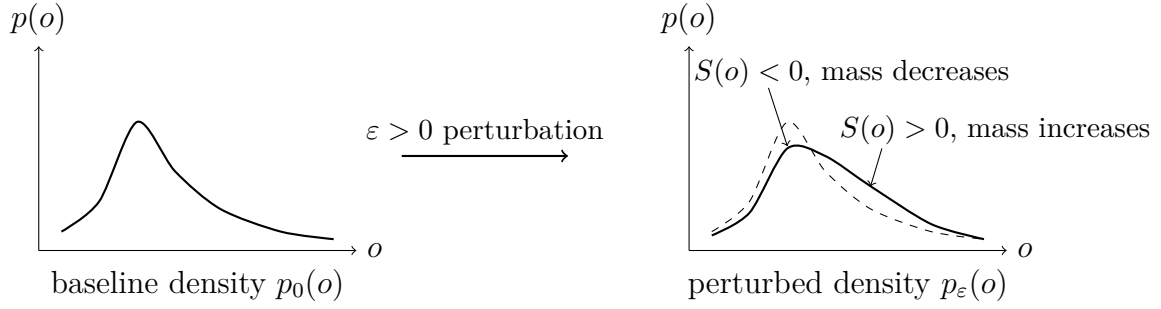

Within the pointwise setting considered above, substituting
$\dot p(o)=S(o)p_0(o)$ into the first-order expansion gives
\[
p_\varepsilon(o)
=
p_0(o)\{1+\varepsilon S(o)\}
+
\mathrm{o}(\varepsilon) \quad\text{pointwise in }o.
\]
This pointwise representation provides a useful interpretation of the score. For positive
$\varepsilon$, locations in the sample space with $S(o)>0$ gain probability mass to first
order, whereas locations with $S(o)<0$ lose probability mass to first order. Thus, the
score describes the local pattern by which probability mass is redistributed along the
path. Reversing the sign of $\varepsilon$ reverses the direction of movement; see
Figure~\ref{fig:path-score} for an illustration.

It is important to distinguish the observation $o$ from the perturbation index $\varepsilon$. The score is not a derivative with respect to $o$, nor does it describe the shape of the density as a function of the observation. Rather, for each fixed observation $o$, the score measures how the probability assigned to that observation changes as the distribution moves along the path $P_\varepsilon$. In this sense, the score describes movement \emph{between distributions}, not movement \emph{within a distribution}. It plays the role for probability distributions that a velocity vector plays for curves in Euclidean space.  

At first sight it may seem surprising that statistical directions are represented by functions rather than finite-dimensional vectors. In ordinary calculus, a direction vector specifies how each coordinate changes. A statistical perturbation must instead describe how probability mass changes across the entire sample space. The resulting direction is therefore naturally represented by a function of the observation rather than by a finite-dimensional vector.

\subsection{From scores to tangent spaces} 

A single path through $P_0$ generates a single score, and hence a single infinitesimal direction of movement. A statistical model, however, typically admits many different paths through the same distribution. Each path generates its own score. To understand the local geometry of the model, we must collect all score directions that can arise from paths remaining inside the model.

This construction is the statistical analogue of forming a tangent plane in ordinary geometry. On a smooth surface embedded in $\mathbb R^3$, many curves can pass through the same point, and each curve has a velocity vector at that point. The tangent plane is obtained by collecting all such velocity vectors. Similarly, a statistical model $\mathcal M$ may contain many paths through $P_0$, each generating its own score direction. The tangent space is obtained by combining all score directions arising from such paths, much as a tangent plane is obtained by combining all velocity vectors arising from curves through a point. 

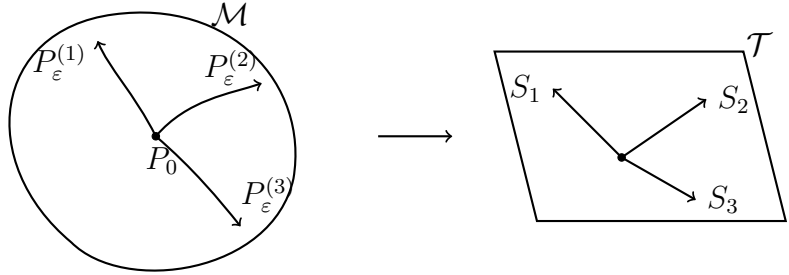
\begin{figure}[t]
\centering
\begin{tikzpicture}[scale=0.7]

\draw[thick] (0,0) .. controls (-1.8,1.5) and (-1.5,4.0) .. (0.8,4.3)
              .. controls (3.0,4.6) and (4.5,3.0) .. (4.2,1.2)
              .. controls (3.8,-0.6) and (1.0,-1.0) .. (0,0);

\node at (3.,4.3) {$\mathcal M$};

\filldraw (1.6,2.0) circle (2pt);
\node[below right] at (1.2,2.0) {$P_0$};

\draw[thick, ->] (1.6,2.0) .. controls (1.2,2.8) and (0.7,3.3) .. (0.5,3.8);
\node[left] at (0.5,3.4) {$P_\varepsilon^{(1)}$};

\draw[thick, ->] (1.6,2.0) .. controls (2.2,2.7) and (3.0,2.8) .. (3.6,3.0);
\node[right] at (2.3,3.3) {$P_\varepsilon^{(2)}$};

\draw[thick, ->] (1.6,2.0) .. controls (2.3,1.4) and (2.8,0.8) .. (3.2,0.3);
\node[right] at (3.0,0.9) {$P_\varepsilon^{(3)}$};

\draw[->, thick] (5.8,2.0) -- (7.2,2.0);
\node[above] at (6.5,2.0) {};

\draw[thick] (8.8,0.4) -- (13.5,0.4) -- (12.7,3.6) -- (8.,3.6) -- cycle;
\node at (13.0,3.75) {$\mathcal T$};

\filldraw (10.4,1.6) circle (2pt);

\draw[->, thick] (10.4,1.6) -- (9.1,2.9) node[left] {$S_1$};
\draw[->, thick] (10.4,1.6) -- (12.0,2.7) node[right] {$S_2$};
\draw[->, thick] (10.4,1.6) -- (11.8,0.8) node[right] {$S_3$};

\end{tikzpicture}
\vspace{-0.25cm}
\caption{Each regular path through $P_0$ generates a score. The tangent space $\mathcal T$ is obtained by collecting all scores arising from paths that remain inside the statistical model $\mathcal M$; that is, $\mathcal T = \overline{\mathrm{span}} \{ \text{scores of regular paths in } \mathcal M\} \subseteq {L^2_0(P_0)}$. }
\label{fig:paths-to-tangent-space}
\end{figure}

Formally, each regular path $\{P_\varepsilon\}\subseteq\mathcal M$ through $P_0$
generates a score $S\in L_0^2(P_0)$. The tangent space is obtained by taking the
closed linear span of all score directions generated by regular paths that remain inside
the statistical model. Informally, the tangent space therefore collects all first-order
directions of movement attainable under the model, together with their linear combinations
and limits; see Figure~\ref{fig:paths-to-tangent-space} for an illustration.


\begin{definition}
The tangent space of the statistical model $\mathcal M$ at $P_0$, denoted $\mathcal T$, is the closed linear span in $L_0^2(P_0)$ of all scores arising from regular paths through $P_0$:
\[
\mathcal T
=
\overline{\mathrm{span}}
\Big\{
S:
S
=
\frac{\partial}{\partial\varepsilon}
\log p_\varepsilon
\Big|_{\varepsilon=0}
\text{ for some regular path } P_\varepsilon\subseteq\mathcal M
\Big\}.
\]
\end{definition} 

The tangent space therefore answers the question: what infinitesimal directions of movement are allowed by the statistical model? 
The \textit{closure} ensures that the tangent space contains not only finite linear combinations
of scores arising from regular paths, but also their $L^2(P_0)$ limits. Thus, a direction
that can be approximated arbitrarily well by linear combinations of attainable score
directions is included in the local linear geometry of the model. This technical condition
also ensures that the geometric projection operations introduced later are well defined.

Strictly speaking, the tangent space is a local object attached to the reference distribution $P_0$, and should be written $\mathcal T(P_0)$. This is exactly analogous to ordinary differential geometry, where each point on a manifold has its own tangent space. Throughout the remainder of the paper, we suppress this dependence and write $\mathcal T$ whenever the reference distribution is clear from context. The same convention will apply to nuisance tangent spaces introduced later. 

This definition also explains why the tangent space depends only on the model. Its construction uses the model $\mathcal M$, the true distribution $P_0$, and the regular paths through $P_0$ that remain inside $\mathcal M$. No parameter of interest appears in the definition. The tangent space describes what movements are possible; it does not yet describe which movements matter for a particular inferential target. That second question is introduced only after a parameter has been specified, and leads to the nuisance tangent space in Section~\ref{sec:nuisance_directions}.

Tangent-space calculations usually involve two complementary steps. First, one constructs regular paths to show that certain score directions are attainable, or can be approximated by linear combinations of attainable score directions. Second, one shows that an arbitrary regular path in the model cannot generate a score outside the proposed space. Thus, if $\mathcal T^\star$ denotes a candidate tangent space, the goal is to verify that the closed linear span of attainable scores is exactly $\mathcal T^\star$. In simple parametric models, this is often immediate because any path through $P_0$ corresponds to a differentiable curve in the finite-dimensional parameter indexing the model. Different paths may move through the parameter space at different speeds, but their scores must lie in the span of the usual parametric score vectors.

\begin{example}
Consider first the Bernoulli model $O\sim\mathrm{Bernoulli}(p)$. Let $P_0=P_{p_0}$. Any regular path through $P_0$ in this model corresponds locally to a differentiable curve $\varepsilon\mapsto p_\varepsilon$ with $p_{\varepsilon=0}=p_0$ and $0<p_\varepsilon<1$ for $\varepsilon$ near zero. Writing $\dot p_0 = \left.\frac{d}{d\varepsilon}p_\varepsilon\right|_{\varepsilon=0}$, the Bernoulli density along the path is $p_\varepsilon(o)=p_\varepsilon^o(1-p_\varepsilon)^{1-o}$. Differentiating the log-density gives $S(o)=\dot p_0\left\{\frac{o}{p_0} - \frac{1-o}{1-p_0}\right\} = \dot p_0\frac{o-p_0}{p_0(1-p_0)}$. Thus every score arising from any regular path in the Bernoulli model is proportional to $o-p_0$. Conversely, the linear path $p_\varepsilon=p_0+\varepsilon a$ generates this direction, with arbitrary scalar multiple determined by $a$. Hence $\mathcal T=\mathrm{span}\{O-p_0\}$. Since every mean-zero function of a Bernoulli random variable is proportional to $O-p_0$, it follows that $\mathcal T=L_0^2(P_0)$. Now consider the Normal location model $\mathcal M=\{N(\mu,1):\mu\in\mathbb R\}$. Any regular path through $P_0=N(\mu_0,1)$ corresponds to a differentiable curve $\varepsilon\mapsto \mu_\varepsilon$ with $\mu_{\varepsilon=0}=\mu_0$. Writing $\dot\mu_0 = \left.\frac{d}{d\varepsilon}\mu_\varepsilon\right|_{\varepsilon=0}$, the score is $S(o) = \dot\mu_0(o-\mu_0)$. Thus every score arising from any regular path in the Normal location model is proportional to $o-\mu_0$. Conversely, the linear path $\mu_\varepsilon=\mu_0+\varepsilon a$ generates any scalar multiple of this direction. Hence $\mathcal T=\mathrm{span}\{O-\mu_0\}$. Unlike the Bernoulli case, $L_0^2(P_0)$ is much larger than this tangent space. For example, $(O-\mu_0)^2-1$ and $(O-\mu_0)^3$ belong to $L_0^2(P_0)$, but they are not proportional to $O-\mu_0$. Therefore they do not belong to $\mathcal T$, and $\mathcal T\subsetneq L_0^2(P_0)$. 
\end{example}

These examples illustrate two possibilities. In some models, the tangent space coincides with the entire space of perturbation directions. In others, the model permits movement in only a small subset of directions. The tangent space therefore captures precisely which infinitesimal perturbations are allowed by the statistical model.

\section{Parameters and Nuisance Directions} 
\label{sec:nuisance_directions} 

In the previous section, we characterized the local geometry of a statistical model through its tangent space. The tangent space describes all infinitesimal directions in which the data-generating distribution may move while remaining inside the model. At that stage, however, no parameter of interest had been specified. The tangent space therefore tells us what perturbations are possible, but not which perturbations matter for a particular inferential problem.

To answer that question, we must introduce the parameter. Once a parameter has been specified, some perturbations change the parameter while others leave it unchanged. This distinction introduces a second layer of geometry: the model determines which directions are possible, whereas the parameter determines which directions are relevant. A simple finite-dimensional analogy in Euclidean space is useful. Suppose $f(x_1,x_2)=x_1$. At a point $x=(x_1,x_2)$, movement in the $x_1$-direction changes the value of $f$, whereas movement in the $x_2$-direction leaves $f$ unchanged. Both directions may be allowed by the Euclidean space, but only one is informative for the function $f$. Similarly, in a statistical model, many perturbations of the distribution may be allowed, but only some change the parameter of interest to first order. 

The central question of this section is therefore: \emph{Among all perturbations allowed by the model, which ones affect the parameter to first order?}

\subsection{How parameters respond to perturbations}

Before introducing the statistical construction, recall the corresponding idea from multivariable calculus. Given a differentiable function $f:\mathbb R^d\to\mathbb R$, the tangent space at a point describes the directions in which movement is possible. Once a function $f$ has been specified, however, one can ask a more refined question: how does $f$ change when moving in a particular direction? This question is answered by the directional derivative. Some directions produce a nonzero directional derivative and therefore change the value of the function to first order, whereas others produce a directional derivative of zero and leave the function unchanged to first order. 

Semiparametric efficiency theory asks the same question for statistical parameters defined on spaces of probability distributions. Let $\psi:\mathcal M \to \mathbb R$ denote a parameter of interest. Given a path $\{P_\varepsilon:\varepsilon\in(-\delta,\delta)\} \subseteq \mathcal M$, the parameter traces out an ordinary real-valued curve
\[
\varepsilon
\mapsto
\psi(P_\varepsilon).
\]

This observation is fundamental. Once a path has been specified, the parameter becomes an ordinary real-valued function of $\varepsilon$. Consequently, ordinary differential calculus applies. The derivative
\[
\left.
\frac{d}{d\varepsilon}
\psi(P_\varepsilon)
\right|_{\varepsilon=0}
\]

\vspace{0.3cm} \noindent 
measures the rate at which the parameter changes as the distribution moves along the path. This quantity is called the \emph{pathwise derivative} of the parameter. Pathwise differentiability may therefore be viewed as the infinite-dimensional analogue of directional differentiability, with regular paths providing the admissible directions of perturbation.\footnote{Pathwise differentiability is closely related to directional (e.g., Gâteaux) differentiability for functionals on infinite-dimensional spaces. The key difference is that probability distributions do not form a linear vector space, so perturbations are represented by regular paths rather than vector addition.}

For a path generating score direction $S$, we denote the pathwise derivative as $\dot\psi(S)=\left.\frac{d}{d\varepsilon}\psi(P_\varepsilon)\right|_{\varepsilon=0}$. For pathwise differentiable parameters, this quantity depends only on the score direction $S$ and not on the particular path used to generate it. This is directly analogous to ordinary directional derivatives, which depend only on the velocity vector of a curve at a point and not on the particular curve used to generate that velocity.

\begin{figure}[t]
\centering
\begin{tikzpicture}[scale=1.0]

\draw[thick, rounded corners=8pt]
(-6,-2.2) rectangle (6.5,2.2);

\node[anchor=north east] at (0.5,2.75) {$\mathcal M$};

\filldraw (0,0) circle (2.3pt);
\node[below] at (0,-0.08) {$P_0$};

\draw[thick, ->]
(0,0) .. controls (-0.7,-0.3) and (-1.5,0.0) .. (-2.6,0.0);

\draw[thick, ->]
(0,0) .. controls (0.7,0.3) and (1.5,-0.1) .. (2.8,-0.2);

\draw[thick, ->]
(0,0) .. controls (-0.7,0.55) and (-1.5,0.8) .. (-2.6,0.95);

\draw[thick, ->]
(0,0) .. controls (-0.6,-0.45) and (-1.55,-0.7) .. (-2.8,-0.85);

\draw[thick, ->]
(0,0) .. controls (0.75,0.7) and (1.65,1.0) .. (2.85,1.25);

\draw[thick, ->]
(0,0) .. controls (0.75,-0.55) and (1.65,-0.75) .. (2.85,-1.05);

\node[align=center] at (3.95, 0.2)
{\scriptsize nuisance path: $\dot\psi(S)=0$,  $S \in \mathcal T_\eta$};

\node[align=center] at (3.5,-1.45)
{\scriptsize informative path: $\dot\psi(S)\neq0$,  $S \in \mathcal T \setminus \mathcal T_\eta $};

\node[align=center] at (3.5,1.55)
{\scriptsize informative path: $\dot\psi(S)\neq 0$, \scriptsize  $S \in \mathcal T \setminus \mathcal T_\eta $} ;

\node[align=center] at (-3.4,0.3)
{\scriptsize $\psi(P_\varepsilon)$ unchanged to first order};

\node[align=center] at (-3.0,1.25)
{\scriptsize $\psi(P_\varepsilon)$ changes to first order};

\node[align=center] at (-3.0,-1.25)
{\scriptsize $\psi(P_\varepsilon)$ changes to first order};

\end{tikzpicture}
\caption{All paths remain inside the same statistical model $\mathcal M$ and therefore generate scores in the model tangent space $\mathcal T$. Paths along which $\psi(P_\varepsilon)$ is unchanged to first order generate nuisance scores; these scores form $\mathcal T_\eta$. Paths along which $\psi(P_\varepsilon)$ changes to first order generate directions that are relevant to the parameter.}
\label{fig:nuisance-paths}
\end{figure}
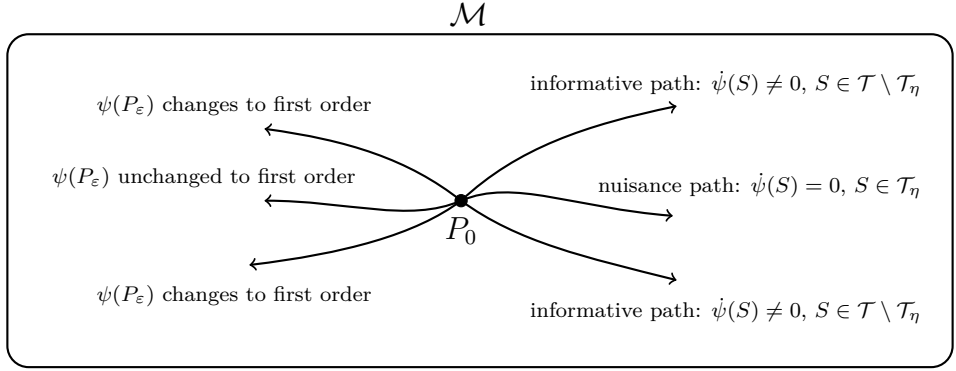

Not every allowable perturbation necessarily affects the parameter. If $\left.\frac{d}{d\varepsilon}\psi(P_\varepsilon)\right|_{\varepsilon=0}\neq 0$, i.e., $\dot\psi(S) \neq 0$, then movement along the path changes the parameter to first order. The parameter is therefore sensitive to that direction. In contrast, if $\left.\frac{d}{d\varepsilon} \psi(P_\varepsilon)\right|_{\varepsilon=0} = 0$, i.e., $\dot\psi(S) = 0$, then the parameter remains unchanged to first order even though the distribution itself moves. Such perturbations are irrelevant to the parameter to first order; see Figure~\ref{fig:nuisance-paths} for illustrations. 

This distinction introduces a second geometric question. The tangent space tells us what movements are possible. The parameter determines which of those movements are relevant. Thus, the same statistical model may contain many allowable perturbations, only some of which matter for the inferential target to first order.

\subsection{The nuisance tangent space}

The preceding discussion suggests a natural classification of directions. Some perturbations change the parameter to first order and are therefore informative. Others leave the parameter unchanged to first order and are therefore uninformative. The collection of all such uninformative directions forms the nuisance tangent space. 

\begin{definition} 
Let $\dot\psi(S) \coloneqq \left.\frac{d}{d\varepsilon}\psi(P_\varepsilon)\right|_{\varepsilon=0}$ denote the pathwise derivative of parameter $\psi(P_0)$ associated with a perturbation direction $S$. The nuisance tangent space associated with parameter $\psi(P_0)$ is defined as
\[
\mathcal T_\eta
=
\{
S\in\mathcal T :
\dot\psi(S) =0
\}.
\]
\end{definition}

As with the tangent space, we suppress the dependence of $\mathcal T_\eta$ on $P_0$ whenever the reference distribution is clear.

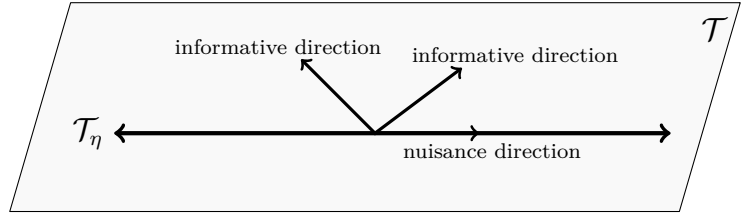
\begin{figure}[t]
\centering

\begin{tikzpicture}[scale=1.15, line cap=round, line join=round]

\filldraw[fill=gray!5, draw=black]
(-4.2,-0.9) -- (3.5,-0.9) -- (4.2,1.5) -- (-3.5,1.5) -- cycle;

\node at (3.9,1.2) {$\mathcal T$};

\coordinate (O) at (0,0);

\draw[ultra thick,->] (O) -- (3.4,0);
\draw[ultra thick,->] (O) -- (-3.0,0);
\node[below] at (-3.3,0.30) {$\mathcal T_\eta$};

\draw[very thick,->] (O) -- (-0.85,0.85);
\node[left] at (0.2,1.) {\scriptsize informative direction};

\draw[very thick,->] (O) -- (1.0,0.75);
\node[right] at (0.3,0.9) {\scriptsize informative direction};

\draw[very thick,->] (O) -- (1.2,0.0);
\node[right] at (0.2,-0.2) {\scriptsize nuisance direction};

\end{tikzpicture}
\caption{The model tangent space $\mathcal T$ contains all allowable infinitesimal perturbations. The nuisance tangent space $\mathcal T_\eta \subseteq \mathcal T$ contains those allowable perturbations along which the parameter is unchanged to first order. Directions in $\mathcal T\setminus \mathcal T_\eta$ affect the parameter to first order.}
\label{fig:nuisance-subspace}
\end{figure}

The nuisance tangent space is therefore the collection of directions that leave the parameter unchanged to first order; see Figure~\ref{fig:nuisance-subspace} for illustration. Because $\mathcal T_\eta$ is defined through the derivative of the parameter, it depends on the inferential target $\psi(.)$, not merely on the statistical model. Geometrically, the directions in $\mathcal T_\eta$ are invisible to the parameter to first order. Movement along a nuisance direction changes the data-generating distribution but does not change the inferential target. 

This interpretation reveals an important distinction. The tangent space is determined entirely by the statistical model and describes all allowable directions of movement. The nuisance tangent space is determined jointly by the model and the parameter, and identifies which of those directions are uninformative for the inferential target.

The dependence of nuisance directions on the parameter can be illustrated using the Normal location-scale model.  

\begin{example}
Consider the Normal location-scale model $\mathcal M = \{N(\mu,\sigma^2):\mu\in\mathbb R,\ \sigma>0\}$. Let $P_0=N(\mu_0,\sigma_0^2)$. A location perturbation may be constructed by varying the mean while holding the variance fixed:
$\mu_\varepsilon = \mu_0+\varepsilon a$ and $\sigma_\varepsilon^2 = \sigma_0^2$. Differentiating the log-density gives the score $S_\mu(o) = a\, \frac{o-\mu_0}{\sigma_0^2}$, which is proportional to $o-\mu_0$. Similarly, a scale perturbation may be constructed by varying the variance while holding the mean fixed: $\mu_\varepsilon = \mu_0$ and $\sigma_\varepsilon^2 = \sigma_0^2+\varepsilon b$. Differentiating the log-density gives $S_{\sigma^2}(o) = \frac{b}{2\sigma_0^4}\left\{(o-\mu_0)^2-\sigma_0^2\right\}$, which is proportional to $(o-\mu_0)^2-\sigma_0^2$. Hence the tangent space is $\mathcal T = \mathrm{span}\big\{O-\mu_0,\, (O-\mu_0)^2-\sigma_0^2\big\}$. Thus, the model admits two infinitesimal directions of movement: a location direction and a scale direction. The score direction $O-\mu_0$ corresponds to perturbations of the location parameter, whereas $(O-\mu_0)^2-\sigma_0^2$ corresponds to perturbations of the scale parameter. Now consider two different parameters defined on exactly the same statistical model. First, consider the mean, $\psi_1(P) = E_P\{O\} = \mu$. Perturbations of the location parameter change the mean, whereas perturbations of the scale parameter do not. Consequently, the nuisance tangent space is $\mathcal T_{\eta,1} = \mathrm{span}\left\{(O-\mu_0)^2-\sigma_0^2\right\}$. The location direction is informative, while the scale direction leaves the parameter unchanged to first order. Next, consider the variance, $\psi_2(P) = \mathrm{Var}_P\{O\} = \sigma^2$. Now the situation is reversed. Perturbations of the scale parameter change the variance, whereas perturbations of the location parameter do not. Hence $\mathcal T_{\eta,2} = \mathrm{span} \left\{ O-\mu_0\right\}$. The location direction is nuisance, while the scale direction is informative. 
\end{example}

These examples use exactly the same statistical model and therefore exactly the same tangent space, but different parameters. They illustrate that whether a direction is nuisance is not a property of the direction alone, but of the direction relative to the inferential target. More generally, the statistical model determines what perturbations are possible and therefore determines the tangent space, whereas the parameter determines which perturbations are relevant and therefore determines the nuisance tangent space.

Having identified the directions that are relevant for a parameter, the next question is how to quantify the sensitivity of the parameter to movement in such directions. 

\section{Influence Functions as Gradients}
\label{sec:IFs_gradients}

In the previous sections, we developed the local geometry of statistical models. We introduced paths, scores, tangent spaces, and nuisance tangent spaces, and interpreted these objects as analogues of curves and directions in ordinary geometry. Having identified the directions in which a distribution can move and distinguished those that are informative from those that are not, we now face a new question: \emph{How should we quantify the sensitivity of a parameter to movement in an informative direction?} In ordinary multivariable calculus, this role is played by the gradient. In semiparametric efficiency theory, it is played by the influence function. 


\subsection{Gradients as representations of derivatives}

Consider a differentiable function $f:\mathbb R^d\to\mathbb R$. Suppose we wish to understand how the value of $f$ changes near a point $x$. Given a direction $v\in\mathbb R^d$, the directional derivative of $f$ in that direction is
\[
D f_x(v)
=
\left.
\frac{d}{d\varepsilon}
f(x+\varepsilon v)
\right|_{\varepsilon=0}.
\]

\vspace{0.3cm}
A derivative is fundamentally not a vector but a rule assigning a rate of change to every direction. The directional derivative evaluates that rule in a particular direction, while the derivative itself is the map $D f_x: \mathbb R^d \to \mathbb R$ that sends each direction $v$ to the corresponding rate of change $D f_x(v)$. For a differentiable function, this map is linear:
\[
D f_x(a v_1+b v_2)
=
aD f_x(v_1)
+
bD f_x(v_2).
\]
Thus, the derivative is a linear functional on the space of directions.

In multivariable calculus, the gradient is introduced as the vector of partial derivatives, $\nabla f(x) = \left(\frac{\partial f}{\partial x_1}(x), \frac{\partial f}{\partial x_2}(x), \ldots, \frac{\partial f}{\partial x_d}(x)\right)^\top$. A remarkable fact from multivariable calculus is that every linear derivative map can be represented by a single vector. Specifically, there exists a unique vector $\nabla f(x)\in\mathbb R^d$ such that for every direction $v$,  
\[
D f_x(v)
=
\nabla f(x)^\top v.
\]

The gradient therefore summarizes all directional derivatives simultaneously. Rather than evaluating the derivative separately in every direction, it suffices to know a single object, namely $\nabla f(x)$. From this perspective, the gradient is not the derivative itself. It is a representation of the derivative map. This observation is the key to understanding influence functions. 

\subsection{The statistical derivatives}

Now consider a parameter $\psi:\mathcal M\to\mathbb R$. As discussed in the previous section, every score direction $S\in\mathcal T$ may be generated by a regular path $\{P_\varepsilon\}$ through the true distribution $P_0$. The sensitivity of the parameter to movement along that path is measured by the pathwise derivative $\dot\psi(S) = \left.\frac{d}{d\varepsilon}\psi(P_\varepsilon)\right|_{\varepsilon=0}$. 

\vspace{0.2cm}
The analogy with ordinary calculus is immediate. Just as the directional derivative measures the sensitivity of a function to movement in direction $v$, the pathwise derivative measures the sensitivity of a parameter to movement in score direction $S$. Moreover, the pathwise derivative assigns a real number to every allowable direction. Consequently,
\[
\dot\psi:
\mathcal T
\to
\mathbb R
\]
may be viewed as a map from directions to rates of change. This is the analogue of the derivative map $D f_x: \mathbb R^d \to \mathbb R$. Notice that the derivative map is defined only on directions that are attainable under the statistical model, namely those belonging to the tangent space $\mathcal T$.

The correspondence is $v \longleftrightarrow S$ and  $D f_x(v) \longleftrightarrow \dot\psi(S)$. Thus, the fundamental object is not a particular directional derivative but the entire derivative map $\dot\psi: \mathcal T \to \mathbb R$.  This map describes how the parameter responds to every infinitesimal perturbation allowed by the statistical model.

Under standard regularity conditions, the pathwise derivative is linear:
\[
\dot\psi(aS_1+bS_2)
=
a\dot\psi(S_1)
+
b\dot\psi(S_2).
\]
Therefore, just like the ordinary derivative, the statistical derivative is a linear functional on a space of directions.

This raises a natural question: If gradients represent derivatives in ordinary calculus, is there an analogous object that represents the statistical derivatives? To answer this question, we must first identify the geometric space in which score directions live.

\subsection{Influence functions as gradients}

Recall from Section~\ref{sec:directions_tangent_space} that every regular score satisfies $E_{P_0}\{S(O)\}=0$ and $E_{P_0}\{S(O)^2\}<\infty$ and belongs to the space $L_0^2(P_0) = \{ f: E_{P_0}\{f(O)\}=0, \, E_{P_0}\{f(O)^2\}<\infty \}$. This space is equipped with the inner product $\langle f,g\rangle = E_{P_0}\{f(O)g(O)\}$. The importance of an inner product is geometric. It provides notions of length, angle, orthogonality, and projection. More fundamentally, it provides a way to represent linear maps through inner products.

This is exactly what happens in ordinary multivariable calculus. The derivative map $Df_x:\mathbb R^d\to\mathbb R$ is represented by a gradient through the relation $Df_x(v) = \nabla f(x)^\top v$. 

The same question therefore arises in semiparametric efficiency theory. Can the statistical derivative map $\dot\psi:\mathcal T\to\mathbb R$ be represented by a single object? 

Pathwise differentiability implies that the derivative map $\dot\psi$ is \textit{linear}, and under standard regularity conditions it is \textit{continuous}. Because the tangent space inherits the inner-product geometry of $L_0^2(P_0)$, a fundamental result from Hilbert-space geometry (the Riesz representation theorem) implies that such maps (continuous linear functional on $\mathcal T$) can be represented as an inner product with an element of $L_0^2(P_0)$. 

\begin{definition}
An influence function\footnote{In the semiparametric literature, the term influence function is also used in connection with asymptotically linear estimators. For regular estimators, these notions coincide. Here we adopt the geometric viewpoint and define influence functions through their role as representations of the pathwise derivative map.} for a pathwise differentiable parameter $\psi(P_0)$ is any function $\varphi\in L_0^2(P_0)$ such that for every score direction $S\in\mathcal T$:
\[
\dot\psi(S)
=
\langle\varphi,S\rangle
=
E_{P_0}
\{\varphi(O)S(O)\}.
\]
\mbox{}
\end{definition} 

The influence function is therefore not the derivative itself. Rather, it is the function that represents the derivative map through the inner product structure of $L_0^2(P_0)$. This is the analogue of the gradient representation $D f_x(v)=\nabla f(x)^\top v$. In ordinary calculus, the gradient represents the derivative map; in semiparametric efficiency theory, the influence function represents the pathwise derivative map. The remaining question is whether this gradient representation is unique. As we shall see next, many different functions can represent the same derivative map. 

The role of the larger space $L_0^2(P_0)$ becomes clearer by considering two special cases. If the tangent space coincides with the entire space, $\mathcal T=L_0^2(P_0)$, then the representation is unique. Any two functions that produce the same inner products with every score direction must be identical. This is precisely what happens in the nonparametric model, where every mean-zero square-integrable perturbation direction is allowed. 

To understand what happens when the tangent space is smaller than the $L_0^2(P_0)$ space, it is helpful to return briefly to ordinary geometry. Suppose allowable directions are restricted to the plane $\mathcal T = \{(v_1,v_2,0):v_1,v_2\in\mathbb R\} \subset \mathbb R^3$. Consider the derivative map $D(v) = v_1+2v_2$. This derivative may be represented by the vector $g=(1,2,0)$, since $D(v)=g^\top v$ for every $v\in\mathcal T$. However, the vector $\tilde g=(1,2,5)$ produces exactly the same directional derivatives on $\mathcal T$, because $\tilde g^\top v = (1,2,5)\cdot(v_1,v_2,0) = v_1+2v_2$. More generally, any vector of the form $(1,2,c)$ represents the same derivative map on $\mathcal T$. The ambiguity arises because the normal direction $(0,0,1)$ is orthogonal to every allowable direction in the plane and therefore cannot be detected through directional derivatives restricted to $\mathcal T$.

The situation is analogous when the tangent space is a proper subspace of $L_0^2(P_0)$. In that case, there exist functions that are orthogonal to every score direction in $\mathcal T$. Adding such a function to $\varphi$ does not change any inner product $\langle \varphi,S\rangle, \, S\in\mathcal T$, and therefore does not change the derivative map. Consequently, multiple functions may represent the same pathwise derivative.

This nonuniqueness will play a central role in the next section. When the tangent space is smaller than $L_0^2(P_0)$, different functions can represent the same derivative map, and we must determine which representative is most useful. Understanding this nonuniqueness leads naturally to orthogonality, projection, and the efficient influence function. 

\section{Orthogonality, Projections, and EIFs}
\label{sec:EIF}

The previous section showed that influence functions play the same mathematical role in semiparametric efficiency theory that gradients play in ordinary calculus. Both are representations of derivative maps. In ordinary calculus, $D f_x(v) = \nabla f(x)^\top v$, while in semiparametric efficiency theory, $\dot\psi(S) = E_{P_0}\{\varphi(O)S(O)\}$. 

At first sight, the analogy appears complete. However, an important subtlety remains. When derivatives are defined on an entire Euclidean space, the gradient is uniquely determined by the derivative map. This uniqueness follows because the derivative is represented on all possible directions of movement. If movement is restricted to a lower-dimensional subspace, however, the same derivative map may admit multiple gradient representations that differ by components orthogonal to the allowable directions. Semiparametric efficiency theory is intrinsically of this latter type. The derivative map is defined only on score directions belonging to the tangent space $\mathcal T$, which may be a proper subspace of $L_0^2(P_0)$. Consequently, influence functions are generally not unique. Understanding why influence functions are not unique leads directly to the central ideas of orthogonality, projection, and efficiency.

\subsection{Gradients under constraints}

\begin{figure}[t]
\centering

\begin{tikzpicture}[scale=1.1]

\filldraw[fill=gray!12,draw=black]
(-2,-0.8)--(2,-0.8)--(3,0.8)--(-1,0.8)--cycle;

\node at (3.9,0.0) {tangent plane};

\coordinate (O) at (0,0);

\draw[->,very thick]
(O)--(1.2,2.0);

\node[right] at (0.4,1.8)
{\small $\nabla f$};

\draw[dashed]
(1.2,2.0)--(1.2,0.4);

\draw[->,very thick]
(O)--(1.2,0.4);

\node[below]
at (1.1,0.25)
{\small $\Pi(\nabla f)$};

\draw[thick]
(1.2,0.4)--(1.2,2.0);

\node[right]
at (1.3,1.2)
{\small normal component};

\end{tikzpicture}

\caption{When movement is constrained to a surface, only the component of the gradient lying in the tangent plane affects directional derivatives. The normal component is irrelevant and is removed by projection.}
\label{fig:projected-gradient}
\end{figure}
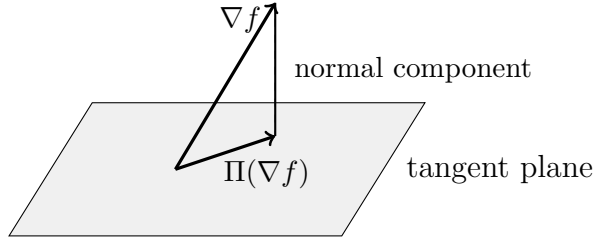

Suppose a smooth surface is embedded in $\mathbb R^3$, and let $f:\mathbb R^3 \to \mathbb R$ be a differentiable function. If movement is restricted to remain on the surface, not every component of the gradient contributes to directional derivatives. Although the gradient $\nabla f$ lives in the full three-dimensional space, only its component tangent to the surface affects directional derivatives along allowable directions. Any component perpendicular to the surface is invisible to motion constrained to remain on the surface. Consequently, the relevant gradient is not the gradient itself but its projection onto the tangent plane. Figure~\ref{fig:projected-gradient} illustrates this idea. 

As a concrete example, suppose movement is restricted to the unit sphere $x_1^2+x_2^2+x_3^2=1$, and consider the function $f(x_1,x_2,x_3)=x_1+x_2+2x_3$. Its gradient is $\nabla f=(1,1,2)$. At the point $x=(0,0,1)$, the tangent plane to the sphere is $\{(v_1,v_2,0):v_1,v_2\in\mathbb R\}$. The gradient decomposes as $\nabla f = (1,1,0) + (0,0,2)$, where the first component lies in the tangent plane and the second component is perpendicular to it. This perpendicular component is often called the \emph{normal component}. Because every allowable direction $v$ lies in the tangent plane, $\nabla f^\top v = (1,1,0)^\top v$. Thus the normal component contributes nothing to directional derivatives along the sphere. When movement is constrained, the relevant gradient is therefore not the full gradient itself but its projection onto the tangent plane.

This geometric observation is the key to understanding efficient influence functions (a.k.a., canonical gradients). Semiparametric efficiency theory faces the same problem. The statistical model determines which perturbation directions are possible. Any component of a gradient that cannot be detected by those perturbations is irrelevant for the derivative. Understanding which components are relevant therefore becomes a problem of orthogonal projection.

\subsection{Influence functions are not necessarily unique}

Recall that an influence function is a function $\varphi \in L_0^2(P_0)$ satisfying $\dot\psi(S) = E_{P_0}\{\varphi(O) S(O)\}$ for every score direction $S\in\mathcal T$. The important point is that this equality is required only for directions belonging to the tangent space.

Now consider a function $h \in \mathcal T^\perp$,  where
\[
\mathcal T^\perp
=
\left\{
h\in L_0^2(P_0)
:
E_{P_0}\{h(O)S(O)\}=0
\text{ for all }
S\in\mathcal T
\right\}.
\]
Since $h$ is orthogonal to every allowable perturbation direction, $E_{P_0}\{h(O)S(O)\}=0$ for all $S\in\mathcal T$. Therefore, for any $h \in \mathcal T^\perp$: 
\[
E_{P_0}\{(\varphi(O)+h(O))S(O)\}
=
E_{P_0}\{\varphi(O) S(O)\}
+
E_{P_0}\{h(O)S(O)\}
=
E_{P_0}\{\varphi(O) S(O)\}. 
\]
It follows that both $\varphi$ and $\varphi+h$ generate exactly the same derivative map, $\dot\psi(S) = E_{P_0}\{(\varphi(O)+h(O))S(O)\} = E_{P_0}\{\varphi(O) S(O)\}$. Thus, influence functions are generally not unique. Rather, they form an entire affine class, 
\[
\text{Class all influence functions}
=  
\varphi+\mathcal T^\perp.
\]

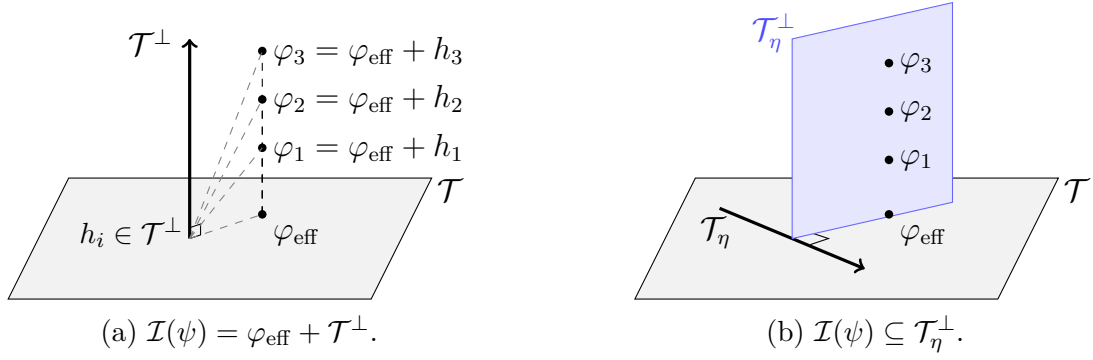
\begin{figure}[t]
\centering

\begin{subfigure}[t]{0.48\textwidth}
\centering

\begin{tikzpicture}[scale=.8]

\filldraw[
fill=gray!10,
draw=black
]
(-3,-1)
--
(3,-1)
--
(4,1)
--
(-2,1)
--
cycle;

\node at (4.3,0.8)
{$\mathcal T$};

\coordinate (O) at (0,0);

\draw[->,very thick]
(O)--(0,3.3);

\node[left]
at (-0.1,3.25)
{$\mathcal T^\perp$};

\draw
(0.0,0)
--
(0.18,0.06)
--
(0.18,0.24)
--
(0,0.18);

\coordinate (phieff) at (1.2,.4);
\fill (phieff) circle (2pt);

\node[below right]
at (phieff)
{$\varphi_{\rm eff}$};

\coordinate (phi1) at (1.2,1.5);
\coordinate (phi2) at (1.2,2.3);
\coordinate (phi3) at (1.2,3.1);
\coordinate (h_points) at (-2.,0.1);

\fill (phi1) circle (2pt);
\fill (phi2) circle (2pt);
\fill (phi3) circle (2pt);

\node[right] at (phi1)
{$\varphi_1=\varphi_{\rm eff}+h_1$};

\node[right] at (phi2)
{$\varphi_2=\varphi_{\rm eff}+h_2$};

\node[right] at (phi3)
{$\varphi_3=\varphi_{\rm eff}+h_3$};

\node[right] at (h_points)
{\small $h_i \in \mathcal T^\perp$};

\draw[dashed] (phieff)--(phi1);
\draw[dashed] (phieff)--(phi2);
\draw[dashed] (phieff)--(phi3);

\draw[dashed,gray]
(0,.0)--(phieff);

\draw[dashed,gray]
(0,0)--(phi1);

\draw[dashed,gray]
(0,0)--(phi2);

\draw[dashed,gray]
(0,0)--(phi3);

\end{tikzpicture}

\caption{$\mathcal I(\psi) = \varphi_{\rm eff} + \mathcal T^\perp$.}
\end{subfigure}
\hfill
\begin{subfigure}[t]{0.48\textwidth}
\centering

\begin{tikzpicture}[scale=.8]

\filldraw[
fill=gray!10,
draw=black
]
(-3,-1)
--
(3,-1)
--
(4,1)
--
(-2,1)
--
cycle;

\node at (4.3,.8)
{$\mathcal T$};

\draw[very thick,->]
(-1.6,0.5)
--
(.8,-0.5);

\node[right]
at (-2.1,0.1)
{$\mathcal T_\eta$};

\draw
(-0.4,0.0)
--
(-0.1,0.1)
--
(0.2,0.0)
--
(-0.1,-0.1);

\draw[
fill=blue!10,
draw=blue!60
]
(-0.4,-0.0)
--
(2.25,0.6)
--
(2.25,3.9)
--
(-0.4,3.3)
--
cycle;

\node[blue!70]
at (-0.7,3.4)
{$\mathcal T_\eta^\perp$};

\coordinate (phieff) at (1.2,.4);
\fill (phieff) circle (2pt);

\node[below right]
at (phieff)
{$\varphi_{\rm eff}$};

\coordinate (phi1) at (1.2,1.3);
\coordinate (phi2) at (1.2,2.1);
\coordinate (phi3) at (1.2,2.9);

\fill (phi1) circle (2pt);
\fill (phi2) circle (2pt);
\fill (phi3) circle (2pt);

\node[right] at (phi1)
{$\varphi_1$};

\node[right] at (phi2)
{$\varphi_2$};

\node[right] at (phi3)
{$\varphi_3$};

\end{tikzpicture}

\caption{$\mathcal I(\psi) \subseteq \mathcal T_\eta^\perp$.}
\end{subfigure}

\caption{Two equivalent geometric characterizations of the class of influence functions. (a) Influence functions differ only by components in $\mathcal T^\perp$, which are orthogonal to every score direction in the tangent space. (b) Every influence function belongs to $\mathcal T_\eta^\perp$, the orthogonal complement of the nuisance tangent space. The efficient influence function is the unique element $\varphi_{\mathrm{eff}}\in \mathcal I(\psi)\cap\mathcal T$. }
\label{fig:IF-class}
\end{figure}

Figure~\ref{fig:IF-class}(a) illustrates this geometry. Adding a component in $\mathcal T^\perp$ changes the influence function only in directions that are orthogonal to all allowable perturbations of the model. Because such components have zero inner product with every score direction, they do not alter the derivative map. This is precisely analogous to adding a normal component to a gradient on a constraint surface. 

A second characterization emerges from the nuisance tangent space. Recall that $\mathcal T_\eta = \{S\in\mathcal T:\dot\psi(S)=0\}$. Every nuisance direction leaves the parameter unchanged to first order. Therefore every influence function satisfies $E_{P_0}\{\varphi(O) S(O)\}=0, \, \forall S\in\mathcal T_\eta$. Hence every influence function belongs to $\mathcal T_\eta^\perp$. Thus, $\mathcal I(\psi) \subseteq \mathcal T_\eta^\perp$. The converse is not generally true: a function may be orthogonal to all nuisance directions without representing the correct derivative on the informative directions. The influence-function class is therefore an affine subset of $\mathcal T_\eta^\perp$, consisting of those functions that agree with the derivative map on all of $\mathcal T$, 
\[
\text{Class all influence functions}
\subseteq 
\mathcal T_\eta^\perp.
\]
\noindent 
Figure~\ref{fig:IF-class}(b) illustrates this geometry. 

\begin{definition}
The influence-function class associated with a pathwise differentiable parameter $\psi(P_0)$, denoted by $\mathcal I(\psi)$, is the collection

\vspace{-0.5cm}
\[
\mathcal I(\psi)
=
\left\{
\varphi\in L_0^2(P_0):
\dot\psi(S)
=
E_{P_0}\{\varphi(O) S(O)\},
\ \forall S\in\mathcal T
\right\}.
\]

\vspace{0.2cm} \noindent
That is, $\mathcal I(\psi)$ contains all influence functions representing the derivative map. We have that $\mathcal I(\psi) = \varphi+\mathcal T^\perp \subseteq \mathcal T_\eta^\perp$.
\end{definition} 

The two characterizations highlight two fundamentally different notions of irrelevance.
The \textit{first} notion of irrelevance is model-based. The orthogonal complement $\mathcal T^\perp$ consists of functions that are orthogonal to every allowable perturbation direction in the model. Consequently, adding a component in $\mathcal T^\perp$ does not change the derivative map because such components have zero inner product with every score direction in the tangent space.
The \textit{second} notion of irrelevance is parameter-based. The nuisance tangent space $\mathcal T_\eta = \{S\in\mathcal T:\dot\psi(S)=0\}$ contains perturbations that are allowable under the model but do not affect the parameter to first order. Together, these two notions of irrelevance induce a natural decomposition of the Hilbert space.

Because $\mathcal T_\eta \subseteq \mathcal T$, the tangent space admits the orthogonal decomposition
$\mathcal T =
\mathcal T_\eta
\oplus
(
\mathcal T
\cap
\mathcal T_\eta^\perp
)$. The first component consists of nuisance directions, while the second consists of directions orthogonal to all nuisance directions. Since $S\in\mathcal T_\eta$ implies $\dot\psi(S)=0$, nuisance directions carry no first-order information about the parameter. Consequently, all first-order information available under the model must lie in $\mathcal T\cap\mathcal T_\eta^\perp$. Combining $L_0^2(P_0) = \mathcal T \oplus \mathcal T^\perp$ with $\mathcal T = \mathcal T_\eta \oplus (\mathcal T\cap\mathcal T_\eta^\perp)$ yields
\[
L_0^2(P_0)
=
\mathcal T^\perp
\oplus
\mathcal T_\eta
\oplus
\left(
\mathcal T
\cap
\mathcal T_\eta^\perp
\right).
\]

\noindent 
The distinctions may be summarized as follows: 
\[
\begin{array}{lll}
\mathcal T^\perp
&:&
\text{functions orthogonal to every allowable perturbation direction},
\\[2mm]
\mathcal T_\eta
&:&
\text{allowable perturbations that do not affect the parameter},
\\[2mm]
\mathcal T
\cap
\mathcal T_\eta^\perp
&:&
\text{allowable perturbations that are informative for the parameter.}
\end{array}
\]

\vspace{0.6cm} 
The efficient influence function emerges by removing both forms of irrelevance simultaneously: it is the unique member of the class of influence functions that lies entirely within the informative component $\mathcal T \cap \mathcal T_\eta^\perp$. 

\subsection{Efficient influence functions as projected gradients}

The previous subsection showed that a pathwise differentiable parameter generally admits an entire class of influence functions. Any two influence functions represent the same derivative map and differ only by a component belonging to $\mathcal T^\perp$. This raises a natural question: if many influence functions represent the same derivative map, is there a preferred representative?


The answer follows from the geometric decomposition above. Since influence functions differ only by components in $\mathcal T^\perp$, the natural representative is obtained by projecting onto $\mathcal T$. The result is the unique influence function in $\mathcal T \cap \mathcal T_\eta^\perp$, the component of the tangent space informative for the parameter.

\begin{definition}
The efficient influence function, also called the canonical gradient, is the unique influence function belonging to $\mathcal T \cap \mathcal T_\eta^\perp$.  Equivalently,
$
\varphi_{\mathrm{eff}}
=
\Pi
\left(
\varphi
\mid
\mathcal T
\cap
\mathcal T_\eta^\perp
\right),
$
where $\varphi$ is any element of $\mathcal I(\psi)$. Since every influence function already belongs to $\mathcal T_\eta^\perp$, projection onto $\mathcal T$ alone yields the same result:
$
\varphi_{\mathrm{eff}}
=
\Pi
\left(
\varphi
\mid
\mathcal T \,
\right).
$
\end{definition}

The efficient influence function is therefore the unique influence function that lies entirely within the informative component of the model geometry. It is obtained by removing the component orthogonal to all allowable perturbation directions and retaining only the component that contributes to the derivative map.

In ordinary geometry, gradients are projected onto allowable directions of movement. In semiparametric efficiency theory, influence functions are projected onto the informative component of the tangent space. Efficient influence functions are therefore projected gradients. The projection interpretation is geometric, but it also has statistical consequences. Different influence functions can represent the same derivative map, yet the efficient influence function is distinguished by the fact that it removes every component that is invisible to the model or irrelevant to the parameter. As we will see in the next section, this projected gradient not only provides the most natural geometric representation of the derivative map, but also plays a central role in the construction of efficient estimators and the characterization of efficiency bounds.

\section{From Geometry to Statistical Inference}
\label{sec:estimation}

The preceding sections developed the geometric foundations of semiparametric efficiency theory. We interpreted influence functions as gradients of statistical functionals and efficient influence functions as projected gradients. A natural question remains: \emph{Why are influence functions so useful for estimation?}

The answer is the same reason gradients are useful in ordinary calculus. Gradients provide first-order approximations to functions. Influence functions provide first-order approximations to statistical parameters. These approximations are what make influence functions central to modern semiparametric inference.

\subsection{Influence functions as first-order approximations}

Recall that for a differentiable function $f:\mathbb R^d\rightarrow\mathbb R$, $f(x+\Delta) = f(x) + \nabla f(x)^\top \Delta + o(\|\Delta\|)$. This first-order Taylor expansion shows that the gradient provides a local linear approximation to the change in the function.

Semiparametric efficiency theory admits an analogous expansion. For notational convenience, let $\varphi_P$ denote an influence function for $\psi$ evaluated at a generic distribution $P$, and write $\varphi_0=\varphi_{P_0}$. Under suitable regularity conditions, many pathwise differentiable parameters admit a first-order von Mises expansion around $P_0$:
$
\psi(P)
-
\psi(P_0)
=
(P-P_0)\varphi_0
+
R(P,P_0),
$
where $(P-P_0)\varphi_0 = \int \varphi_0(o)\,d(P-P_0)(o)$, and $R(P,P_0)$ denotes a higher-order remainder term.

This expansion is the statistical analogue of the first-order Taylor expansion. The influence function $\varphi_0$ plays the role of the gradient, the signed measure $P-P_0$ plays the role of the displacement $\Delta$, and the term $(P-P_0)\varphi_0$ is the leading linear approximation to the change in the parameter.

For estimation, however, $\varphi_0$ is unknown because it depends on the unknown distribution $P_0$. Therefore, influence-function-based estimators replace $\varphi_0$ by an estimated influence function $\varphi_{\widehat P}$, obtained by evaluating the same influence-function formula at an estimate $\widehat P$ of $P_0$. This distinction is important: $\varphi_0$ describes the ideal first-order behavior at the truth, whereas $\varphi_{\widehat P}$ is the computable version used to construct estimators.

\subsection{Influence-function based estimators}

Suppose $\widehat P$ is an estimate of the true data-generating distribution $P_0$ and consider the plug-in estimator $\psi(\widehat P)$ of $\psi(P_0)$. Applying the von Mises expansion around $\widehat P$ gives
$\psi(P_0)-\psi(\widehat P) = (P_0-\widehat P)\varphi_{\widehat P} + R(P_0,\widehat P)$. Since $\widehat P\varphi_{\widehat P}=0$, this may be written as
$\psi(\widehat P)-\psi(P_0) = - P_0\varphi_{\widehat P} + R(\widehat P,P_0)$. Thus, to first order, the error of the plug-in estimator is captured by the term $-P_0\varphi_{\widehat P}$. Since $P_0\varphi_{\widehat P}$ is unknown, it is natural to estimate it by
the empirical average $P_n\varphi_{\widehat P}=\frac{1}{n}\sum_{i=1}^n\varphi_{\widehat P}(O_i)$. This leads to the one-step corrected plug-in estimator
$
\widehat\psi_{\mathrm{1step}}
=
\psi(\widehat P)
+
P_n\varphi_{\widehat P}
=
\psi(\widehat P)
+
\frac1n
\sum_{i=1}^n
\varphi_{\widehat P}(O_i).
$

Geometrically, the one-step estimator is obtained by taking a plug-in estimate and adding a gradient correction. The influence function estimates the first-order error induced by replacing $P_0$ with $\widehat P$, and the empirical average supplies the corresponding correction. This construction is analogous to using a first-order Taylor correction in ordinary calculus.

Many modern estimation procedures may be viewed as attempts to exploit this same geometry. One-step estimators explicitly add the influence-function correction to a plug-in estimate. Targeted minimum loss estimators (TMLEs) update an initial estimate so that the empirical mean of the efficient influence function is close to zero. Debiased machine learning and related procedures similarly use efficient influence functions to remove first-order bias induced by flexible machine learning estimators of nuisance quantities.

Although the implementation details differ, these methods share a common geometric foundation. Each uses the efficient influence function to characterize and remove first-order estimation error. From this perspective, influence functions are not merely tools for deriving estimators after the fact. They are the objects that determine how estimators should be constructed in the first place.

\subsection{Efficiency as a projection phenomenon}

Section~\ref{sec:EIF} showed that different influence functions can represent the same derivative map, and that the efficient influence function is the unique representative obtained by projecting onto the informative component of the tangent space. This projection has an important statistical consequence.

Orthogonal projection has a fundamental variational property: among all elements representing the same derivative map, the projection has the smallest norm. Consequently,
$
\|\varphi_{\mathrm{eff}}\|
=
\min_{\widetilde\varphi \, \in \, \mathcal I(\psi)}
\|\widetilde\varphi\|,
$
where $\mathcal I(\psi)$ denotes the class of influence functions representing the derivative map of $\psi$ at $P_0$.

This geometric fact becomes statistically meaningful because the influence functions introduced in the previous sections also determine the asymptotic behavior of regular estimators. Specifically, regular asymptotically linear estimators admit an expansion of the form 
$
\sqrt n
\{\widehat\psi-\psi(P_0)\}
=
\frac{1}{\sqrt n}
\sum_{i=1}^n
\varphi(O_i)
+
o_p(1).
$
A fundamental result in semiparametric theory states that, for regular estimators, the function $\varphi$ appearing in this expansion must belong to $\mathcal I(\psi)$; that is, it must be an influence function representing the derivative map of the parameter $\psi$.

It follows from the central limit theorem that
$
\mathrm{Var}
\left(
\sqrt n
\{\widehat\psi-\psi(P_0)\}
\right)
=
E_{P_0}\{\varphi(O)^2\}
=
\|\phi\|^2.
$
The asymptotic variance of a regular estimator is therefore determined by the squared norm of its influence function. Because the efficient influence function has the smallest norm among all influence functions representing the same derivative map, $ \|\varphi_{\mathrm{eff}}\|^2 = E_{P_0}\{\varphi_{\mathrm{eff}}(O)^2\}$ is the smallest asymptotic variance achievable by a regular estimator. This quantity is known as the semiparametric efficiency bound.


The efficient influence function therefore occupies a unique role in semiparametric theory: geometrically, it is the projected gradient onto the informative model geometry; statistically, it determines the smallest achievable asymptotic variance. Thus, the same object captures the local geometry of the parameter, guides estimator construction, and characterizes the limits of statistical efficiency.

\section{Worked Examples}
\label{sec:examples}

The preceding sections developed semiparametric efficiency theory as differential calculus on a space of probability distributions. Table~\ref{tab:comparisons} in Appendix~\ref{app:sec:notation} summarizes the principal geometric objects and the distinct roles of the statistical model and parameter of interest. This section illustrates these ideas in several familiar settings; detailed pathwise derivations and tangent-space calculations are provided in Appendix~\ref{app:sec:derivations}.


Each example highlights a different aspect of the geometry. The first revisits the average treatment effect in the unrestricted nonparametric model. The second shows that restricting the statistical model need not improve efficiency if the removed directions were already nuisance. The third shows how removing informative directions simplifies the efficient influence function. The fourth illustrates efficient influence functions as projections by comparing an unadjusted and covariate-adjusted gradient under randomized treatment.

Before turning to specific examples, it is useful to make the notion of a path more concrete. A probability distribution may be perturbed in many different ways while remaining inside a statistical model. In practice, exponential tilting is often convenient because it automatically preserves nonnegativity and normalization of probability distributions. Similarly, for binary treatment mechanisms, perturbing the logit of the propensity score automatically preserves the constraint that probabilities remain between zero and one.


The paths constructed below are merely convenient devices for generating score directions. Different regular paths may generate the same direction, while the geometric objects of interest---scores, tangent spaces, pathwise derivatives, and efficient influence functions---do not depend on the particular path representation.

Let $O=(X,A,Y)$, with $A\in\{0,1\}$, and consider the average
treatment effect
$
\psi(P) = E_P\{\mu_1(X)-\mu_0(X)\},
$
where $\mu_a(x) = E_P(Y\mid A=a,X=x)$. 
Assume the observed-data distribution admits the factorization
$
p(o)
=
p(y\mid a,x)\,
p(a\mid x)\,
p(x).
$
A regular path through $P_0$ may be constructed by perturbing one or more of these components:
$
p_\varepsilon(o)
=
p_\varepsilon(y\mid a,x)\,
p_\varepsilon(a\mid x)\,
p_\varepsilon(x).
$

For mixed discrete--continuous covariates $X$, let $\nu_X$ denote a dominating measure for the distribution of $X$, combining counting measure for discrete components and Lebesgue measure for continuous components. Then $p_0(x)$ denotes the density or mass function of $X$ with respect to $\nu_X$. For a mean-zero square-integrable function $s_X(x)$ satisfying  $E_{P_0}\{s_X(X)\}=0$, one convenient local path for the marginal distribution of $X$ is
$
p_{\varepsilon}(x)
=
\frac{p_0(x)\exp\{\varepsilon s_X(x)\}}
{\int p_0(u)\exp\{\varepsilon s_X(u)\}\,d\nu_X(u)},
$
which keeps the density nonnegative and normalized. Its score is 
$
S_X(O) = 
\left.
\frac{\partial}{\partial\varepsilon}
\log p_\varepsilon(X)
\right|_{\varepsilon=0}
=
s_X(X),
$
provided $s_X$ is centered under $P_0$.

For the binary treatment mechanism, write $\pi_0(x)=P_0(A=1\mid X=x)$. A path for the conditional distribution of $A\mid X$ may be defined by perturbing the logit of the propensity score:
$
p_\varepsilon(a\mid x)
=
\pi_\varepsilon(x)^a
\{1-\pi_\varepsilon(x)\}^{1-a}$, where 
$
\operatorname{logit}\pi_\varepsilon(x)
=
\operatorname{logit}\pi_0(x)
+
\varepsilon h_A(x).
$
The corresponding score is
$
S_{A\mid X}(O)
=
\left.
\frac{\partial}{\partial\varepsilon}
\log p_\varepsilon(A\mid X)
\right|_{\varepsilon=0}
=
\{A-\pi_0(X)\}h_A(X).
$
Thus treatment-mechanism directions are functions with conditional mean zero given $X$.

For the outcome regression component, a generic path for the conditional distribution of $Y\mid A,X$ may be written as
$
p_\varepsilon(y\mid a,x)
=
\frac{
p_0(y\mid a,x)\exp\{\varepsilon s_Y(y,a,x)\}
}{
\int p_0(u\mid a,x)\exp\{\varepsilon s_Y(u,a,x)\}\,d\nu_Y(u)
},
$
where $s_Y$ satisfies $E_{P_0}\{s_Y(Y,A,X)\mid A,X\}=0$. The score for this component is
$
S_{Y\mid A,X}(O)
=
\left.
\frac{\partial}{\partial\varepsilon}
\log p_\varepsilon(Y\mid A,X)
\right|_{\varepsilon=0}
=
s_Y(Y,A,X).
$

Putting the pieces together gives a path $p_\varepsilon(o) = p_\varepsilon(y\mid a,x) p_\varepsilon(a\mid x) p_\varepsilon(x)$, with score
$
S(O)
= 
S_X(X)
+
S_{A\mid X}(A,X)
+
S_{Y\mid A,X}(Y,A,X).
$
This decomposition is the concrete version of the tangent-space decomposition
$
\mathcal T
=
\mathcal T_X
\oplus
\mathcal T_{A\mid X}
\oplus
\mathcal T_{Y\mid A,X}.
$
It also shows why tangent-space components correspond to parts of the likelihood factorization. Perturbing $p(x)$ creates a covariate-distribution score; perturbing $p(a\mid x)$ creates a treatment-mechanism score; perturbing $p(y\mid a,x)$ creates an outcome-mechanism score.

\subsection{ATE in the nonparametric model}

Given the path construction from the previous subsection, the ATE under perturbed distribution $P_\epsilon$ becomes 
$
\psi(P_\varepsilon)
=
\iint
y \big\{p_\varepsilon(y \mid A=1, x)- p_\varepsilon(y \mid A=0, x) \big\} \, p_\varepsilon(x) \, d\nu_Y(y) \,d\nu_X(x). 
$
The expression depends on the covariate distribution $p_\varepsilon(x)$ and the conditional outcome distributions $p_\varepsilon(y\mid a,x)$, but it does not depend directly on the treatment mechanism $p_\varepsilon(a\mid x)$. Consequently, treatment-mechanism perturbations do not change the parameter to first order and belong to the nuisance tangent space: $\mathcal T_{A\mid X} \subseteq \mathcal T_\eta$. 

Because the model is completely unrestricted, every mean-zero square-integrable perturbation direction is allowed. Consequently, $\mathcal T = L_0^2(P_0)$. There are therefore no model-imposed constraints on the gradient. The efficient influence function is 
\[
\varphi_{\mathrm{eff}}(O)
=
\frac{A}{\pi_0(X)}
\{Y-\mu_{1,0}(X)\}
-
\frac{1-A}{1-\pi_0(X)}
\{Y-\mu_{0,0}(X)\}
+
\mu_{1,0}(X)-\mu_{0,0}(X)
-
\psi(P_0),
\]
where $\pi_0(X)=P_0(A=1\mid X)$ and $\mu_{a, 0}(X) = E[Y \, | \, A=a, X]$. This decomposition reflects the geometry of the parameter. The term $\frac{A}{\pi_0(X)}\{Y-\mu_{1,0}(X)\}-\frac{1-A}{1-\pi_0(X)}\{Y-\mu_{0,0}(X)\}$ corresponds to perturbations of the conditional outcome distribution, while the term $\mu_{1,0}(X)-\mu_{0,0}(X)-\psi(P_0)$ corresponds to perturbations of the covariate distribution. Treatment-mechanism perturbations are absent because they are nuisance directions for the average treatment effect.

This example illustrates the full geometric pipeline: Paths generate scores, scores generate the tangent space, the parameter induces a derivative map on that tangent space, influence functions represent the derivative map, and the efficient influence function is obtained by projection onto the informative component of the model geometry.  

\subsection{ATE with a known treatment mechanism}

We again consider the average treatment effect, but now under a more restrictive statistical model in which the treatment mechanism is known:
$
P_0(A=1\mid X)=\pi_0(X).
$

At first sight, one might expect this additional information to improve efficiency. Surprisingly, for the average treatment effect, it does not. The reason is geometric: the directions removed by the restriction were already nuisance directions for the parameter.

To see this, recall that the average treatment effect depends on the covariate distribution and outcome mechanism through
\[
\psi(P_\varepsilon)
=
\iint
y
\Big\{
p_\varepsilon(y\mid A=1,x)
-
p_\varepsilon(y\mid A=0,x)
\Big\}
p_\varepsilon(x)
\,d\nu_Y(y)\,d\nu_X(x),
\]
and does not depend directly on the treatment mechanism
$p_\varepsilon(a\mid x)$.
Consequently,
$
\mathcal T_{A\mid X}
\subseteq
\mathcal T_\eta.
$
Treatment-mechanism perturbations change the observed-data distribution but leave the parameter unchanged to first order.

The restriction nevertheless changes the statistical model. Because the treatment mechanism is fixed, admissible paths have the form
$
p_\varepsilon(o)
=
p_\varepsilon(y\mid a,x)
\pi_0(x)^a
\{1-\pi_0(x)\}^{1-a}
p_\varepsilon(x),
$
rather than
$
p_\varepsilon(y\mid a,x)
p_\varepsilon(a\mid x)
p_\varepsilon(x).
$

The corresponding score decomposition becomes
$
S
=
S_X
+
S_{Y\mid A,X},
$
since $S_{A\mid X}=0$. The tangent space therefore shrinks from
$
\mathcal T_X
\oplus
\mathcal T_{A\mid X}
\oplus
\mathcal T_{Y\mid A,X}
$
to
$
\mathcal T
=
\mathcal T_X
\oplus
\mathcal T_{Y\mid A,X}.
$

Although the tangent space is smaller, the informative component of the tangent space is unchanged because the removed directions belonged entirely to the nuisance tangent space. Consequently, the efficient influence function is identical to that obtained in the unrestricted nonparametric model.

This example illustrates an important refinement of the geometric intuition. Restricting the statistical model shrinks the tangent space, but efficiency improves only when the restriction removes directions that contribute to the informative component of the gradient. Removing directions that were already nuisance does not improve the efficiency bound.

\subsection{ATE with a known outcome regression}

We again consider the average treatment effect but now suppose the outcome conditional mean function is known: 
$
\mu_a(x)
=
E_P\{Y\mid A=a,X=x\},
\
a\in\{0,1\}. 
$

Unlike the previous example, this restriction changes the efficient influence function. Because the outcome regression is known, all variation associated with estimating the conditional means disappears. The efficient influence function becomes
$
\varphi_{\mathrm{eff}}(O)
=
\mu_1(X)-\mu_0(X)-\psi(P_0).
$

Comparing this expression with the efficient influence function from the unrestricted nonparametric model,
\[
\frac{A}{\pi_0(X)}
\{Y-\mu_1(X)\}
-
\frac{1-A}{1-\pi_0(X)}
\{Y-\mu_0(X)\}
+
\mu_1(X)-\mu_0(X)-\psi(P_0),
\]
shows that the augmentation terms have disappeared entirely.

The geometric reason is that the outcome-regression restriction removes informative directions from the tangent space. Although the treatment mechanism remains unrestricted, perturbations of the conditional outcome distribution are now required to preserve the known conditional means. The observed-data distribution still factorizes as
$
p(o)
=
p(y\mid a,x)
p(a\mid x)
p(x),
$
but only perturbations satisfying the mean constraints are allowed. Consequently, the tangent space becomes
$
\mathcal T
=
\mathcal T_X
\oplus
\mathcal T_{A\mid X}
\oplus
\mathcal T^{*}_{Y\mid A,X},
$
where
$
\mathcal T^{*}_{Y\mid A,X}
=
\{
S:
E_{P_0}\{S\mid A,X\}=0,
\;
E_{P_0}\{YS\mid A,X\}=0
\}.
$

The space $\mathcal T^{*}_{Y\mid A,X}$ contains perturbations of the conditional outcome distribution that preserve the conditional means. Such perturbations may alter higher moments of $Y\mid A,X$, such as the conditional variance, skewness, or tail behavior, while leaving the outcome regression unchanged.

Because the average treatment effect depends on the outcome distribution only through the conditional means $\mu_1(x)$ and $\mu_0(x)$, every direction in $\mathcal T^{*}_{Y\mid A,X}$ is nuisance. Likewise, perturbations of the treatment mechanism remain nuisance because the parameter does not depend directly on $p(a\mid x)$. Therefore,
$
\mathcal T^{*}_{Y\mid A,X}
\subseteq
\mathcal T_\eta$, and 
$
\mathcal T_{A\mid X}
\subseteq
\mathcal T_\eta.
$

The only remaining informative directions are those that perturb the covariate distribution $p(x)$. Consequently, the efficient influence function consists solely of the covariate component
$\mu_1(X)-\mu_0(X)-\psi(P_0)$. 

This example illustrates the geometric effect of removing informative directions from the tangent space. In the previous example, the known treatment mechanism removed directions that were already nuisance, leaving the efficient influence function unchanged. Here, the outcome-regression restriction removes directions that contribute information about the parameter. As a result, the efficient influence function simplifies substantially and the efficiency bound decreases.

\subsection{ATE under randomized treatment}

We now consider a model in which treatment is randomized with respect to covariates:
$
A \perp X.
$
Under this restriction, the observed-data density factorizes as
$
p(o)
=
p(y\mid a,x)\,
p(a)\,
p(x).
$
Let
$
\pi_0
=
P_0(A=1),
\ 
\mu_a(x)
=
E_{P_0}\{Y\mid A=a,X=x\},
\
\theta_a
=
E_{P_0}\{Y\mid A=a\}.
$
Because treatment is randomized, the average treatment effect may be written in two equivalent ways:
$
\psi(P_0)
=
E_{P_0}\{\mu_1(X)-\mu_0(X)\}
=
\theta_1 
-
\theta_0
$
The second representation suggests the influence function
$
\varphi_{\rm unadj}(O)
=
\frac{A}{\pi_0}
\{Y-\theta_1\}
-
\frac{1-A}{1-\pi_0}
\{Y-\theta_0\}.
$

This influence function corresponds to the usual unadjusted difference-in-means estimator. It is a valid representative of the derivative map, but it is not the efficient influence function. To understand why, we examine the geometry induced by the randomization restriction.

Because treatment is independent of the covariates, regular paths must preserve the factorization
$
p_\varepsilon(o)
=
p_\varepsilon(y\, | \, a,x)\,
p_\varepsilon(a)\,
p_\varepsilon(x).
$
The corresponding score decomposition is
$
S
=
S_{Y\mid A,X}
+
S_A
+
S_X,
$
where
$
E_{P_0}\{S_{Y|A,X}\mid A,X\}
=
0,
\
E_{P_0}\{S_A\}
=
0,
\
E_{P_0}\{S_X\}
=
0.
$
The tangent space is therefore
$
\mathcal T
=
\mathcal T_{Y\mid A,X}
\oplus
\mathcal T_A
\oplus
\mathcal T_X.
$

This differs from the nonparametric model because treatment-mechanism directions of the form
$
(A-\pi(X))h(X)
$
are no longer allowed. Such perturbations would induce dependence between $A$ and $X$ and therefore violate the randomization restriction.

The key observation is that the unadjusted influence function does not belong to this tangent space. Using
$
Y-\theta_a
=
(Y-\mu_a(X))
+
(\mu_a(X)-\theta_a),
$
we may write
\[
\varphi_{\rm unadj}
=
\frac{A}{\pi_0}
\{Y-\mu_1(X)\}
-
\frac{1-A}{1-\pi_0}
\{Y-\mu_0(X)\}
+
R(A,X),
\]
where
$
R(A,X)
=
\frac{A}{\pi_0}
\{\mu_1(X)-\theta_1\}
-
\frac{1-A}{1-\pi_0}
\{\mu_0(X)-\theta_0\}.
$
The remainder term contains components of the form
$
(A-\pi_0)
\{\mu_1(X)-\mu_0(X)\},
$
which do not belong to
$
\mathcal T_A
\oplus
\mathcal T_X.
$
Hence part of $\varphi_{\rm unadj}$ lies outside the tangent space induced by the randomized-treatment model.

The efficient influence function is obtained by projecting $\varphi_{\rm unadj}$ onto the tangent space:
$
\varphi_{\rm eff}
=
\Pi(\varphi_{\rm unadj}\mid \mathcal T).
$
This projection removes the components orthogonal to $\mathcal T$ and yields
\[
\varphi_{\rm eff}(O)
=
\frac{A}{\pi_0}
\{Y-\mu_1(X)\}
-
\frac{1-A}{1-\pi_0}
\{Y-\mu_0(X)\}
+
\mu_1(X)-\mu_0(X)
-
\psi(P_0).
\]
This is the familiar covariate-adjusted influence function. It exploits the fact that, under randomization, the treatment groups share the same covariate distribution.

The relationship between the two influence functions is geometric. Their difference is
\[
\varphi_{\rm unadj}(O)
-
\varphi_{\rm eff}(O)
=
\frac{A}{\pi_0}
\{\mu_1(X)-\theta_1\}
-
\frac{1-A}{1-\pi_0}
\{\mu_0(X)-\theta_0\}
-
\{\mu_1(X)-\mu_0(X)-\psi(P_0)\}.
\]
This difference belongs to $\mathcal T^\perp$. Consequently,
$
\varphi_{\rm unadj}
=
\varphi_{\rm eff}
+
h,
\
h\in\mathcal T^\perp.
$
Both functions represent the same derivative map, but only $\varphi_{\rm eff}$ lies entirely in the tangent space. 

This example brings the projection interpretation full circle. The unadjusted and adjusted influence functions represent the same derivative map and differ only by a component that is invisible to all allowable perturbations under the randomized-treatment model. The efficient influence function is obtained by projecting away this irrelevant component. Thus the familiar efficiency gain from covariate adjustment in randomized experiments is a direct consequence of the projection geometry.

\section{Conclusion}
\label{sec:conclusion}


We began with a simple question: how do we perform differential calculus when the object being differentiated is a probability distribution rather than a point in Euclidean space? All of the constructions developed in this tutorial are local: they describe the geometry of a statistical model in an infinitesimal neighborhood of a reference distribution $P_0$.

The answer led naturally to the central objects of semiparametric efficiency theory. A probability distribution plays the role of a point. A path of distributions plays the role of a curve. A score plays the role of a velocity vector. A tangent space collects the allowable directions of movement. A nuisance tangent space collects the allowable directions that do not change the parameter. An influence function represents the derivative map and therefore plays the role of a gradient. An efficient influence function is the projected gradient obtained by removing components that are invisible to the statistical model and retaining only the component that lies in the informative part of the tangent space.

The present tutorial has focused exclusively on first-order geometry. Just as ordinary calculus extends beyond gradients to higher-order derivatives, semiparametric efficiency theory also admits higher-order notions of differentiation. These ideas lead to higher-order influence functions, projected score methods, and higher-order asymptotic approximations \citep{waterman1996projected,van2014higher}. Exploring these extensions would require a richer geometric framework and lies beyond the scope of this tutorial.

The central geometric distinction is therefore between two roles: the statistical model determines what infinitesimal movements are possible, while the parameter determines which of those movements are informative. Semiparametric efficiency theory studies the interaction between these two structures. The efficient influence function is the object that remains after projecting the parameter gradient onto the directions that are both allowed by the model and relevant for the parameter.

Viewed in this way, semiparametric efficiency theory is not a collection of isolated technical constructions. It is a coherent geometric theory. Influence functions are gradients, efficient influence functions are projected gradients, and efficiency bounds are squared lengths of those projected gradients. Semiparametric efficiency theory is, at its core, differential calculus on a space of probability distributions.



\vspace{1cm}

\begingroup
\renewcommand{\baselinestretch}{1}
\selectfont  
\setlength{\bibsep}{12pt}    
\bibliographystyle{abbrvnat}
\bibliography{references}

\begin{thebibliography}{22}
\providecommand{\natexlab}[1]{#1}
\providecommand{\url}[1]{\texttt{#1}}
\expandafter\ifx\csname urlstyle\endcsname\relax
  \providecommand{\doi}[1]{doi: #1}\else
  \providecommand{\doi}{doi: \begingroup \urlstyle{rm}\Url}\fi

\bibitem[Begun et~al.(1983)Begun, Hall, Huang, and
  Wellner]{begun1983information}
J.~M. Begun, W.~J. Hall, W.-M. Huang, and J.~A. Wellner.
\newblock Information and asymptotic efficiency in parametric-nonparametric
  models.
\newblock \emph{The Annals of Statistics}, 11\penalty0 (2):\penalty0 432--452,
  1983.

\bibitem[Bickel et~al.(1993)Bickel, Klaassen, Bickel, Ritov, Klaassen, Wellner,
  and Ritov]{bickel1993efficient}
P.~J. Bickel, C.~A. Klaassen, P.~J. Bickel, Y.~Ritov, J.~Klaassen, J.~A.
  Wellner, and Y.~Ritov.
\newblock \emph{Efficient and adaptive estimation for semiparametric models},
  volume~4.
\newblock Springer, 1993.

\bibitem[Chamberlain(1987)]{chamberlain1987asymptotic}
G.~Chamberlain.
\newblock Asymptotic efficiency in estimation with conditional moment
  restrictions.
\newblock \emph{Journal of econometrics}, 34\penalty0 (3):\penalty0 305--334,
  1987.

\bibitem[Chernozhukov et~al.(2018)]{chernozhukov2018double}
V.~Chernozhukov et~al.
\newblock Double/debiased machine learning for treatment and structural
  parameters.
\newblock \emph{The Econometrics Journal}, 21:\penalty0 C1--C68, 2018.

\bibitem[Hampel(1974)]{hampel1974influence}
F.~R. Hampel.
\newblock The influence curve and its role in robust estimation.
\newblock \emph{Journal of the american statistical association}, 69\penalty0
  (346):\penalty0 383--393, 1974.

\bibitem[Hines et~al.(2022)Hines, Dukes, Diaz-Ordaz, and
  Vansteelandt]{hines2022demystifying}
O.~Hines, O.~Dukes, K.~Diaz-Ordaz, and S.~Vansteelandt.
\newblock Demystifying statistical learning based on efficient influence
  functions.
\newblock \emph{The American Statistician}, 76\penalty0 (3):\penalty0 292--304,
  2022.

\bibitem[Kennedy(2016)]{kennedy2016semiparametric}
E.~H. Kennedy.
\newblock Semiparametric theory and empirical processes in causal inference.
\newblock In \emph{Statistical causal inferences and their applications in
  public health research}, pages 141--167. Springer, 2016.

\bibitem[Kosorok(2008)]{kosorok2008introduction}
M.~R. Kosorok.
\newblock \emph{Introduction to empirical processes and semiparametric
  inference}.
\newblock Springer, 2008.

\bibitem[Le~Cam(1986)]{lecam1986asymptotic}
L.~Le~Cam.
\newblock \emph{Asymptotic Methods in Statistical Decision Theory}.
\newblock Springer, New York, 1986.

\bibitem[Newey(1990)]{newey1990semiparametric}
W.~K. Newey.
\newblock Semiparametric efficiency bounds.
\newblock \emph{Journal of applied econometrics}, 5\penalty0 (2):\penalty0
  99--135, 1990.

\bibitem[Newey(1994)]{newey1994asymptotic}
W.~K. Newey.
\newblock The asymptotic variance of semiparametric estimators.
\newblock \emph{Econometrica: Journal of the Econometric Society}, pages
  1349--1382, 1994.

\bibitem[Pfanzagl(1982)]{pfanzagl1982general}
J.~Pfanzagl.
\newblock \emph{Contributions to a General Asymptotic Statistical Theory},
  volume~13 of \emph{Lecture Notes in Statistics}.
\newblock Springer, New York, 1982.

\bibitem[Robins and Rotnitzky(1995)]{robins1995semiparametric}
J.~M. Robins and A.~Rotnitzky.
\newblock Semiparametric efficiency in multivariate regression models with
  missing data.
\newblock \emph{Journal of the American Statistical Association}, 90\penalty0
  (429):\penalty0 122--129, 1995.

\bibitem[Robins et~al.(1994)Robins, Rotnitzky, and Zhao]{robins1994estimation}
J.~M. Robins, A.~Rotnitzky, and L.~P. Zhao.
\newblock Estimation of regression coefficients when some regressors are not
  always observed.
\newblock \emph{Journal of the American statistical Association}, 89\penalty0
  (427):\penalty0 846--866, 1994.

\bibitem[Scharfstein et~al.(1999)Scharfstein, Rotnitzky, and
  Robins]{scharfstein1999adjusting}
D.~O. Scharfstein, A.~Rotnitzky, and J.~M. Robins.
\newblock Adjusting for nonignorable drop-out using semiparametric nonresponse
  models.
\newblock \emph{Journal of the American Statistical Association}, 94\penalty0
  (448):\penalty0 1096--1120, 1999.

\bibitem[Tsiatis(2006)]{tsiatis2006semiparametric}
A.~A. Tsiatis.
\newblock \emph{Semiparametric theory and missing data}.
\newblock Springer, 2006.

\bibitem[van~der Laan and Robins(2003)]{laan2003unified}
M.~J. van~der Laan and J.~M. Robins.
\newblock \emph{Unified methods for censored longitudinal data and causality}.
\newblock Springer, 2003.

\bibitem[van~der Laan and Rose(2011)]{van2011targeted}
M.~J. van~der Laan and S.~Rose.
\newblock \emph{Targeted learning: causal inference for observational and
  experimental data}, volume~4.
\newblock Springer, 2011.

\bibitem[van~der Vaart(1991)]{van1991differentiable}
A.~van~der Vaart.
\newblock On differentiable functionals.
\newblock \emph{The Annals of Statistics}, pages 178--204, 1991.

\bibitem[van~der Vaart(2014)]{van2014higher}
A.~van~der Vaart.
\newblock Higher order tangent spaces and influence functions.
\newblock \emph{Statistical Science}, pages 679--686, 2014.

\bibitem[van~der Vaart(2000)]{van2000asymptotic}
A.~W. van~der Vaart.
\newblock \emph{Asymptotic statistics}, volume~3.
\newblock Cambridge university press, 2000.

\bibitem[Waterman and Lindsay(1996)]{waterman1996projected}
R.~P. Waterman and B.~G. Lindsay.
\newblock Projected score methods for approximating conditional scores.
\newblock \emph{Biometrika}, 83\penalty0 (1):\penalty0 1--13, 1996.

\end{thebibliography}
\endgroup

\clearpage
\appendix

\setcounter{section}{0}
\renewcommand{\thesection}{S\arabic{section}}
\renewcommand{\thesubsection}{S\arabic{section}.\arabic{subsection}}
\renewcommand{\thesubsubsection}{S\arabic{section}.\arabic{subsection}.\arabic{subsubsection}}

\setcounter{figure}{0}
\renewcommand{\thefigure}{S\arabic{figure}}

\setcounter{table}{0}
\renewcommand{\thetable}{S\arabic{table}}

\setcounter{equation}{0}
\renewcommand{\theequation}{S\arabic{equation}}

\section{Conceptual and notational reference}
\label{app:sec:notation}

\begin{table}[h]
\centering
\caption{Glossary of notation.}
\label{tab:notation}
\setlength{\extrarowheight}{5pt}
\begin{tabular}{ll}
\hline
\textbf{Symbol} & \textbf{Meaning}
\\[0.08cm] \hline 
$O$ & Observed data vector
\\
$\mathcal O$ & Sample space of $O$
\\
$o$ & Realization of $O$
\\
$P$ & Generic probability distribution
\\
$P_0$ & True data-generating distribution
\\
$\mathcal M$ & Statistical model (collection of distributions)
\\
$P_\varepsilon$ & Regular path through $P_0$
\\
$p_\varepsilon$ & Density corresponding to $P_\varepsilon$
\\
$\psi(P)$ & Statistical parameter (functional)
\\
$S$ & Score function associated with a path
\\
$L_0^2(P_0)$ & Mean-zero square-integrable functions under $P_0$
\\
$\mathcal T$ & Tangent space of the model at $P_0$
\\
$\dot\psi(S)$ & Pathwise derivative in direction $S$
\\
$\mathcal T_\eta$ & Nuisance tangent space
\\
$\varphi$ & Influence function
\\
$\varphi_{\mathrm{eff}}$ & Efficient influence function (canonical gradient)
\\
$\mathcal I(\psi)$ & The class of all influence functions for $\psi$
\\
$\mathcal T^\perp$ & Orthogonal complement of $\mathcal T$
\\
$\mathcal T_\eta^\perp$ & Orthogonal complement of $\mathcal T_\eta$
\\
$\Pi(\cdot\mid V)$ & Orthogonal projection onto subspace $V$
\\
$E_P\{\cdot\}$ & Expectation under distribution $P$
\\
$E_{P_0}\{\cdot\}$ & Expectation under the true distribution
\\
$\langle f,g\rangle$ & Inner product defined as $E_{P_0}\{f(O)g(O)\}$
\\
$||f||$ & $L^2(P_0)$ norm of $f$ defined as ${E_{P_0}\{f(O)^2\}}^{1/2}$
\\[0.1cm] \hline
\end{tabular}
\end{table}

\clearpage 

\begin{dictionarybox}
\[
\begin{array}{ccc}
\\[-12.5mm]
\textbf{Multivariable calculus}
&
&
\textbf{Semiparametric efficiency theory}
\\[2mm]
\hline
\\[-1mm]
x, x_0 \in \mathbb R^d
&
\longleftrightarrow
&
P, P_0 \in \mathcal M
\\[3mm]
f:\mathbb R^d \to \mathbb R
&
\longleftrightarrow
&
\psi:\mathcal M \to \mathbb R
\\[3mm]
f(x), f(x_0)
&
\longleftrightarrow
&
\psi(P), \psi(P_0)
\\[2mm]
x_\epsilon
&
\longleftrightarrow
&
P_\varepsilon
\\[2mm]
\epsilon \mapsto x_\varepsilon
&
\longleftrightarrow
&
\epsilon \mapsto P_\varepsilon
\\[3mm]
\left.\dfrac{d}{d\varepsilon}x_\varepsilon\right|_{\varepsilon=0}
&
\longleftrightarrow
&
\left.\dfrac{\partial}{\partial\varepsilon}\log p_\varepsilon\right|_{\varepsilon=0}
\\[6mm]
\text{first-order description}
&
\longleftrightarrow
&
\text{score function}
\\[2mm]
v
&
\longleftrightarrow
&
S
\\[2mm]
\nabla f(x)
&
\longleftrightarrow
&
\varphi
\\[3mm]
\nabla f(x)^\top v
&
\longleftrightarrow
&
E_{P_0}\{\varphi(O)S(O)\}
\\[2mm]
\text{tangent directions}
&
\longleftrightarrow
&
\mathcal T
\\[2mm]
\text{normal directions}
&
\longleftrightarrow
&
\mathcal T^\perp
\\[2mm]
\text{gradient}
&
\longleftrightarrow
&
\varphi
\\[2mm]
\text{projected gradient}
&
\longleftrightarrow
&
\varphi_{\mathrm{eff}}
\end{array}
\]
\end{dictionarybox}

\begin{table}[h]
\begin{center}
\renewcommand{\arraystretch}{1.5}
\scalebox{0.85}{
\begin{tabular}{>{\raggedright\arraybackslash}p{0.38\linewidth} >{\raggedright\arraybackslash}p{0.45\linewidth} >{\raggedright\arraybackslash}p{0.15\linewidth}}
\toprule
\textbf{Object} & \textbf{Meaning} & \textbf{Depends on} \\
\midrule
$L_0^2(P_0)$: Hilbert space  & All mean-zero square-integrable functions of the observed data & $P_0$ \\
$\calT$: model tangent space  & All first-order perturbations that remain inside the model & $P_0$ and $\calM$ \\
$\calT_\eta$: nuisance tangent space  & Allowed perturbations that leave $\psi(P)$ unchanged to first order & $P_0$, $\calM$, and $\psi$ \\
$\calT^\perp$: orthogonal complement of $\mathcal T$ & Directions excluded by the model restrictions & $P_0$ and $\calM$ \\
$\calT_\eta^\perp$: orthogonal complement of $\mathcal T_\eta$ & Directions orthogonal to nuisance variation & $P_0$, $\calM$, and $\psi$ \\
$\varphi_{\mathrm{eff}}$: efficient influence function & Canonical gradient; part of the gradient that carries information about $\psi$ &  $P_0$, $\calM$, and $\psi$  \\
\bottomrule
\end{tabular}
}
\end{center}
\vspace{-0.35cm}
\caption{Geometric interpretation of the main spaces and functions used in the paper.}
\label{tab:comparisons}
\end{table}

\clearpage
\section{Differentiability in quadratic mean}
\label{app:sec:dqm}

The main text develops scores through the log-density derivative
\[
S(o)
=
\left.
\frac{\partial}{\partial\varepsilon}
\log p_\varepsilon(o)
\right|_{\varepsilon=0},
\]
which provides a direct interpretation of the score as the relative
first-order redistribution of probability mass along a path. This section
briefly describes the standard quadratic-mean formulation underlying smooth
statistical paths and explains its connection to the score construction used
in the main text.

Suppose that the distributions under consideration are dominated by a common
measure $\nu$, and let $p=dP/d\nu$ and $q=dQ/d\nu$. The Hellinger distance
between $P$ and $Q$ may be defined, up to a conventional scaling factor, by
\[
H^2(P,Q)
=
\int
\left\{
\sqrt{p(o)}-\sqrt{q(o)}
\right\}^2
\,d\nu(o).
\]
Thus, Hellinger distance measures proximity between distributions through
the $L^2(\nu)$ distance between their square-root densities. The map
\[
P
\longmapsto
\sqrt p
\]
therefore embeds dominated probability distributions into $L^2(\nu)$.
Moreover, because
\[
\|\sqrt p\|_{L^2(\nu)}^2
=
\int p(o)\,d\nu(o)
=
1,
\]
square-root densities lie on the unit sphere of $L^2(\nu)$, restricted to
its nonnegative part.

Let $\{P_\varepsilon:\varepsilon\in(-\delta,\delta)\}$ be a dominated path
through $P_0$, with densities $p_\varepsilon$. The path is
\emph{differentiable in quadratic mean} at $\varepsilon=0$ if there exists
$S\in L^2(P_0)$ such that
\[
\int
\left[
\sqrt{p_\varepsilon(o)}
-
\sqrt{p_0(o)}
-
\frac{\varepsilon}{2}
S(o)\sqrt{p_0(o)}
\right]^2
d\nu(o)
=
o(\varepsilon^2).
\]
The function $S$ is the score of the path. Equivalently, the square-root
density path admits the first-order expansion
\[
\sqrt{p_\varepsilon}
=
\sqrt{p_0}
+
\frac{\varepsilon}{2}
S\sqrt{p_0}
+
o(\varepsilon)
\qquad
\text{in }L^2(\nu).
\]
Hence the $L^2(\nu)$ velocity of the square-root density path is
\[
\frac{1}{2}S\sqrt{p_0}.
\]
The score $S$ is the corresponding representation of this local direction
in $L^2(P_0)$. This gives a precise interpretation of the statement in the
main text that scores play the role of statistical velocity vectors.

The mean-zero property of the score also has a geometric interpretation in
this representation. Since $\sqrt{p_\varepsilon}$ remains on the unit sphere,
its first-order velocity must be orthogonal to the radial direction
$\sqrt{p_0}$. Therefore,
\[
0
=
\left\langle
\sqrt{p_0},
\frac{1}{2}S\sqrt{p_0}
\right\rangle_{L^2(\nu)}
=
\frac{1}{2}
\int S(o)p_0(o)\,d\nu(o)
=
\frac{1}{2}E_{P_0}\{S(O)\}.
\]
Thus,
\[
E_{P_0}\{S(O)\}=0.
\]
Together with $S\in L^2(P_0)$, this shows that the score belongs to
$L_0^2(P_0)$, the Hilbert space used throughout the main text.

Finally, the quadratic-mean definition agrees with the familiar log-density
score under suitable regularity conditions. Suppose that
$\varepsilon\mapsto p_\varepsilon(o)$ is sufficiently smooth and that the
required interchange and integrability conditions hold. Then
\[
\left.
\frac{\partial}{\partial\varepsilon}
\sqrt{p_\varepsilon(o)}
\right|_{\varepsilon=0}
=
\frac{1}{2}
\frac{\dot p(o)}{\sqrt{p_0(o)}}
=
\frac{1}{2}
\left\{
\frac{\dot p(o)}{p_0(o)}
\right\}
\sqrt{p_0(o)},
\]
where
\[
\dot p(o)
=
\left.
\frac{\partial}{\partial\varepsilon}
p_\varepsilon(o)
\right|_{\varepsilon=0}.
\]
Comparing this expression with the quadratic-mean expansion identifies
\[
S(o)
=
\frac{\dot p(o)}{p_0(o)}
=
\left.
\frac{\partial}{\partial\varepsilon}
\log p_\varepsilon(o)
\right|_{\varepsilon=0}
\]
on the support of $P_0$. Thus, for sufficiently smooth dominated paths, the
log-density derivative used in the main text is the score arising from
quadratic-mean differentiability.

The pointwise construction in the main text and the quadratic-mean
formulation above therefore describe the same local object under standard
regularity conditions. The former emphasizes the redistribution of
probability mass, while the latter provides the Hilbert-space differentiability
structure underlying the local geometry of semiparametric efficiency theory.

\clearpage
\section{Detailed derivations for the worked examples}
\label{app:sec:derivations}

Throughout this section let \(O=(X,A,Y)\), \(A\in\{0,1\}\), and
\[
\psi(P)=E_P\{\mu_1(X)-\mu_0(X)\},
\qquad
\mu_a(x)=E_P(Y\mid A=a,X=x).
\]
Write \(\pi_0(x)=P_0(A=1\mid X=x)\), \(\mu_{a,0}(x)=E_{P_0}(Y\mid A=a,X=x)\), and
\[
m_0(X)=\mu_{1,0}(X)-\mu_{0,0}(X).
\]
Under a regular path \(P_\varepsilon\) through \(P_0\), write the score decomposition as
\[
S(O)=S_X(X)+S_{A\mid X}(A,X)+S_{Y\mid A,X}(Y,A,X),
\]
where
\[
E_0\{S_X(X)\}=0,\qquad
E_0\{S_{A\mid X}(A,X)\mid X\}=0,\qquad
E_0\{S_{Y\mid A,X}(Y,A,X)\mid A,X\}=0.
\]

\subsection*{\underline{ATE in the nonparametric model.}}

In the nonparametric model,
\[
\mathcal T
=
\mathcal T_X\oplus\mathcal T_{A\mid X}\oplus\mathcal T_{Y\mid A,X}
=
L_0^2(P_0).
\]
Along a regular path,
\[
\psi(P_\varepsilon)
=
\int \{\mu_{1,\varepsilon}(x)-\mu_{0,\varepsilon}(x)\}
\,p_\varepsilon(x)\,d\nu_X(x).
\]
Differentiating gives
\[
\dot\psi(S)
=
E_0\big[
\{\mu_{1,0}(X)-\mu_{0,0}(X)-\psi(P_0)\}S_X(X)
\big]
+
E_0\left[
\dot\mu_1(X)-\dot\mu_0(X)
\right].
\]
For the outcome component,
\[
\dot\mu_a(x)
=
\frac{\partial}{\partial\varepsilon}
E_\varepsilon(Y\mid A=a,X=x)
\bigg|_{\varepsilon=0}
=
E_0\{Y S_{Y\mid A,X}(Y,A,X)\mid A=a,X=x\}.
\]
Since \(E_0\{S_{Y\mid A,X}\mid A,X\}=0\),
\[
\dot\mu_a(x)
=
E_0\{(Y-\mu_{a,0}(X))S_{Y\mid A,X}\mid A=a,X=x\}.
\]
Therefore,
\[
E_0\{\dot\mu_1(X)\}
=
E_0\left[
\frac{A}{\pi_0(X)}
\{Y-\mu_{1,0}(X)\}
S_{Y\mid A,X}
\right],
\]
and
\[
E_0\{\dot\mu_0(X)\}
=
E_0\left[
\frac{1-A}{1-\pi_0(X)}
\{Y-\mu_{0,0}(X)\}
S_{Y\mid A,X}
\right].
\]
Thus
\[
\dot\psi(S)
=
E_0\left[
\left\{
\frac{A}{\pi_0(X)}(Y-\mu_{1,0}(X))
-
\frac{1-A}{1-\pi_0(X)}(Y-\mu_{0,0}(X))
+
m_0(X)-\psi(P_0)
\right\}S
\right],
\]
because the displayed function is orthogonal to \(S_{A\mid X}\). Hence
\[
\phi_{\mathrm{eff}}(O)
=
\frac{A}{\pi_0(X)}\{Y-\mu_{1,0}(X)\}
-
\frac{1-A}{1-\pi_0(X)}\{Y-\mu_{0,0}(X)\}
+
m_0(X)-\psi(P_0).
\]

\subsection*{\underline{ATE with a known treatment mechanism.}}

Suppose \(P(A=1\mid X)=\pi_0(X)\) is known. Then admissible paths satisfy
\[
p_\varepsilon(o)
=
p_\varepsilon(y\mid a,x)\,
p_0(a\mid x)\,
p_\varepsilon(x),
\]
so that
\[
S(O)=S_X(X)+S_{Y\mid A,X}(Y,A,X),
\]
and
\[
\mathcal T
=
\mathcal T_X\oplus\mathcal T_{Y\mid A,X}.
\]
The parameter satisfies
\[
\psi(P_\varepsilon)
=
\int
\{\mu_{1,\varepsilon}(x)-\mu_{0,\varepsilon}(x)\}
p_\varepsilon(x)\,d\nu_X(x),
\]
so the same calculation as in the nonparametric model gives
\[
\dot\psi(S)
=
E_0\left[
\left\{
\frac{A}{\pi_0(X)}(Y-\mu_{1,0}(X))
-
\frac{1-A}{1-\pi_0(X)}(Y-\mu_{0,0}(X))
+
m_0(X)-\psi(P_0)
\right\}S
\right].
\]
Therefore the efficient influence function is unchanged:
\[
\phi_{\mathrm{eff}}(O)
=
\frac{A}{\pi_0(X)}\{Y-\mu_{1,0}(X)\}
-
\frac{1-A}{1-\pi_0(X)}\{Y-\mu_{0,0}(X)\}
+
m_0(X)-\psi(P_0).
\]
Equivalently, the known-treatment-mechanism restriction removes
\(\mathcal T_{A\mid X}\), but \(\mathcal T_{A\mid X}\subseteq\mathcal T_\eta\);
hence no informative directions are removed.

\subsection*{\underline{ATE with a known outcome regression.}}

Suppose \(\mu_a(x)=E_P(Y\mid A=a,X=x)\) is known for \(a=0,1\). Then admissible outcome scores must preserve the conditional means:
\[
0
=
\frac{\partial}{\partial\varepsilon}
E_\varepsilon(Y\mid A=a,X=x)
\bigg|_{\varepsilon=0}
=
E_0\{Y S_{Y\mid A,X}\mid A=a,X=x\}.
\]
Together with
\[
E_0\{S_{Y\mid A,X}\mid A,X\}=0,
\]
this gives
\[
\mathcal T_{Y\mid A,X}^*
=
\left\{
S:
E_0(S\mid A,X)=0,\;
E_0(YS\mid A,X)=0
\right\}.
\]
The tangent space is
\[
\mathcal T
=
\mathcal T_X
\oplus
\mathcal T_{A\mid X}
\oplus
\mathcal T_{Y\mid A,X}^*.
\]
Since the conditional means are fixed,
\[
\psi(P_\varepsilon)
=
\int
\{\mu_1(x)-\mu_0(x)\}
p_\varepsilon(x)\,d\nu_X(x),
\]
and therefore
\[
\dot\psi(S)
=
E_0[
\{\mu_1(X)-\mu_0(X)-\psi(P_0)\}S_X(X)
].
\]
The function
\[
\phi_{\mathrm{eff}}(O)
=
\mu_1(X)-\mu_0(X)-\psi(P_0)
\]
belongs to \(\mathcal T_X\), is orthogonal to
\(\mathcal T_{A\mid X}\), and is orthogonal to
\(\mathcal T_{Y\mid A,X}^*\). Hence
\[
\dot\psi(S)=E_0\{\phi_{\mathrm{eff}}(O)S(O)\},
\]
and
\[
\phi_{\mathrm{eff}}(O)
=
\mu_1(X)-\mu_0(X)-\psi(P_0).
\]

\subsection*{\underline{ATE under randomized treatment.}}

Suppose \(A\perp X\). Then
\[
p(o)=p(y\mid a,x)p(a)p(x),
\]
and admissible scores decompose as
\[
S(O)=S_{Y\mid A,X}(Y,A,X)+S_A(A)+S_X(X),
\]
where
\[
E_0\{S_A(A)\}=0,
\qquad
E_0\{S_X(X)\}=0,
\qquad
E_0\{S_{Y\mid A,X}\mid A,X\}=0.
\]
Thus
\[
\mathcal T
=
\mathcal T_{Y\mid A,X}\oplus\mathcal T_A\oplus\mathcal T_X.
\]
Let
\[
\pi_0=P_0(A=1),
\qquad
\theta_a=E_0(Y\mid A=a).
\]
Since \(A\perp X\),
\[
\psi(P_0)=\theta_1-\theta_0
=
E_0\{\mu_{1,0}(X)-\mu_{0,0}(X)\}.
\]
The unadjusted influence function is
\[
\phi_{\mathrm{unadj}}(O)
=
\frac{A}{\pi_0}(Y-\theta_1)
-
\frac{1-A}{1-\pi_0}(Y-\theta_0).
\]
This represents the derivative map, but it need not lie in \(\mathcal T\). Decompose
\[
Y-\theta_a
=
\{Y-\mu_{a,0}(X)\}
+
\{\mu_{a,0}(X)-\theta_a\}.
\]
Then
\[
\phi_{\mathrm{unadj}}(O)
=
\frac{A}{\pi_0}\{Y-\mu_{1,0}(X)\}
-
\frac{1-A}{1-\pi_0}\{Y-\mu_{0,0}(X)\}
+
R(A,X),
\]
where
\[
R(A,X)
=
\frac{A}{\pi_0}\{\mu_{1,0}(X)-\theta_1\}
-
\frac{1-A}{1-\pi_0}\{\mu_{0,0}(X)-\theta_0\}.
\]
The projection of \(R(A,X)\) onto
\(\mathcal T_A\oplus\mathcal T_X\) is
\[
m_0(X)-\psi(P_0).
\]
Indeed, define
\[
h(O)
=
R(A,X)-\{m_0(X)-\psi(P_0)\}.
\]
For any \(S_X\in\mathcal T_X\),
\[
E_0\{h(O)S_X(X)\}
=
E_0[
E_0\{h(O)\mid X\}S_X(X)
]
=0,
\]
because under randomization,
\[
E_0\{R(A,X)\mid X\}
=
m_0(X)-\psi(P_0).
\]
For any \(S_A\in\mathcal T_A\),
\[
E_0\{h(O)S_A(A)\}
=
E_0[
E_0\{h(O)\mid A\}S_A(A)
]
=0,
\]
because
\[
E_0\{h(O)\mid A=1\}
=
\frac{1}{\pi_0}E_0\{\mu_{1,0}(X)-\theta_1\}
-
E_0\{m_0(X)-\psi(P_0)\}
=0,
\]
and similarly
\[
E_0\{h(O)\mid A=0\}=0.
\]
Finally, since \(h\) is a function only of \((A,X)\),
\[
E_0\{h(O)S_{Y\mid A,X}(O)\}=0.
\]
Thus \(h\in\mathcal T^\perp\). Therefore
\[
\Pi(\phi_{\mathrm{unadj}}\mid\mathcal T)
=
\frac{A}{\pi_0}\{Y-\mu_{1,0}(X)\}
-
\frac{1-A}{1-\pi_0}\{Y-\mu_{0,0}(X)\}
+
m_0(X)-\psi(P_0).
\]
Hence
\[
\phi_{\mathrm{eff}}(O)
=
\frac{A}{\pi_0}\{Y-\mu_{1,0}(X)\}
-
\frac{1-A}{1-\pi_0}\{Y-\mu_{0,0}(X)\}
+
\mu_{1,0}(X)-\mu_{0,0}(X)-\psi(P_0).
\]
Moreover,
\[
\phi_{\mathrm{unadj}}-\phi_{\mathrm{eff}}
=
h
\in\mathcal T^\perp,
\]
so the adjusted influence function is the projection of the unadjusted influence function
onto the randomized-treatment tangent space.

\end{document}